\documentstyle[epsf]{mn}

\title[The Formation and Evolution of Clusters of Galaxies]{The Formation and
Evolution of Clusters of Galaxies in Different Cosmogonies}

\author[A.~Huss et al.]{A.~Huss,$^1$, B. Jain,$^1$, M.~Steinmetz,$^{1,2}$ \\
$^1$Max-Planck-Institut f\"ur Astrophysik, Postfach 1523, 85740 Garching,
Germany\\
$^2$Department for Astronomy, University of California, Berkeley, CA 94720,
USA}

\newcommand{\etal}{{et al.} \thinspace}
\newcommand{\ie}{i.e.\thinspace}
\newcommand{\eg}{e.g.\thinspace}
\newcommand{\Msun}{\mbox{M$_{\odot}$}}

\begin{document}

\maketitle

\date{}

\begin{abstract}
The formation of galaxy clusters in hierarchically clustering universes is
investigated by means of high resolution N-body simulations. The
simulations are performed using a newly developed multi-mass scheme
which combines a PM code
with a high resolution N-body code.  Numerical effects due to time stepping and
gravitational softening are investigated as well as the influence of the
simulation box size and of the assumed boundary conditions. Special emphasis is
laid on the formation process and the influence of various cosmological
parameters.  Cosmogonies with massive neutrinos are also considered. 
Differences between clusters in the same
cosmological model seem to dominate over differences due differing
background cosmogony.
The cosmological model can alter the time evolution of cluster
collapse, but the merging pattern remains fairly similar, \eg number
of mergers and mass ratio of mergers. 
The gross properties of a halo, such as its size and
total angular momentum, also evolve in a similar manner for all 
cosmogonies and can be described using analytical models.  It is shown that the
density distribution of a halo shows a characteristic radial dependence which
follows a power law with a slope of $\alpha=-1$ at small and $\alpha=-3$ at
large radii, independent of the background cosmogony or the considered
redshift.  The shape of the density profiles follows the generic form
proposed by Navarro et al. (1996) for
all hierarchically clustering scenarios and retains very little information
about the formation process or the cosmological model. Only the central matter
concentration of a halo is correlated to the formation time and
therefore to the corresponding cosmogony.  We emphasise the role of
non-radial motions of the halo particles in the evolution of
the density profile. 

\end{abstract}


\section{Introduction}

Clusters of galaxies are the most massive, gravitationally bound objects in the
universe. Since their average density is only several hundred times as large as
the critical density of the universe and since their collapse time is comparable
to the age of the universe, they are likely to have formed recently.  Such a
recent formation is also implied by substructure observed for many galaxy
clusters.  It is conceivable that non-linear dynamics and violent relaxation has
not yet erased all information about the initial conditions and the formation
process itself. Therefore, galaxy clusters may give insight into how structure
has grown in the universe. In order to extract these insights from the structure
of a galaxy cluster observed at a given epoch, its formation process has to be
understood in detail. One important aspect is the connection between the actual
structure of dark matter haloes and the features of a given theory of structure
formation. The density profile of a dark matter halo at a given epoch may be
determined by the initial power spectrum as well as by the present density
parameter $\Omega_0$ and/or the value of the cosmological constant $\Lambda$.

Such a dependence between the initial power spectrum and the density profiles of
virialized objects was pointed out by Hoffman \& Shaham (1985) and Hoffman
(1988). Their analytic work was based on the assumption of Gaussian random
fields with scale free power spectra $P(k) \propto k^n$. Based on the secondary
infall model of Gunn \& Gott (1972), and the self-similar solution of Fillmore
\& Goldreich (1984) and Bertschinger (1985), they found that the slope of the
density profiles becomes steeper with increasing spectral index $n$.  Analytic
calculations, however, are based on simplifying assumptions (\eg spherical
symmetry) and, therefore, cannot consider all aspects of gravitational
collapse. N-body simulations, which begin with generic initial conditions,
provide an attractive alternative. Though the effects of gas dynamics are
neglected, they are likely to represent a realistic description of the formation
of galaxy clusters: observations suggest that dark matter dominates the mass of
galaxy clusters and largely determines their gravitational potential.  Only in
the highest density regions near the center of a cluster are the time scales
sufficiently short that the cluster dynamics can be influenced by hydrodynamical
effects.

Within the last couple of years, multi-mass techniques have been developed which
allow one to investigate the formation of individual clusters with high
numerical resolution. Navarro {\it et al.} (1996a) (hereafter NFW) have
investigated the structure of dark matter haloes which form in a cold dark
matter (CDM) universe ($\Omega_0 = 1$). They found that the density profiles of
haloes do not follow a power law, but tend to have a slope $\alpha = \frac{d \ln
\varrho}{d \ln r}$ with $\alpha=-1$ near the cluster center and $\alpha = -3$ at
large radii. Over more the four orders of magnitude in mass, the density
profiles follow a universal law, which can be parameterized by
\begin{equation}
\frac{\varrho(r)}{\varrho_b} = \frac{\delta_n}{\frac{r}{a_n} (1 + \frac{r}{a_n})^2}.
\label{nfw}
\end{equation}
The two free parameter are the scale radius $a_n$ which defines the scale where
the profile shape changes from $\alpha >-2$ to $\alpha <-2$ and the
characteristic overdensity $\delta_n$.
Equation~(\ref{nfw}) differs in its asymptotic  behaviour at large radii from
the
profile
\begin{equation}
\frac{\varrho(r)}{\varrho_b} = \frac{\delta_h}{\frac{r}{a_h} (1 + \frac{r}{a_h})^3}.
\label{her}
\end{equation}
which was proposed by Hernquist (1990) to describe the mass profiles of
elliptical galaxies.  This profile has also been used by Dubinsky \& Carlberg
(1991) to fit the density distribution of haloes which were formed in their
CDM-type simulation. Lacey
\& Cole (1996) extended the work of NFW to scale free power spectra with $n =
-2, -1$, and 0 in an $\Omega_0=1$ cosmogony, though with a lower numerical
resolution.  Tormen {\it et al.} (1996) have done simulations with an $n = -1$
scale free power spectrum and $\Omega=1$ and with an even higher numerical
resolution than NFW. Recently, Navarro, Frenk and White (1996b) extended their
study to cosmogonies with $\Omega < 1$ and with a non-vanishing cosmological
constant $\Lambda$.  Haloes which form in all of these cosmological models seem
to be well described by equation~(\ref{nfw}). The scale radius and the central
overdensity seem to be related to the formation time of the halo (Navarro
\etal 1996b). These results suggest that the density profile found by NFW is
quite generic for any scenario in which structures form due to hierarchical
clustering. The power spectrum and cosmological parameters may only indirectly
enter by specifying the typical formation epoch of a halo of a given
mass. Consequently, they possibly affect the profiles of galaxy clusters only by
specifying the dependence of the characteristic radius $a$ on the total mass of
a cluster.

In this paper we concentrate on the profiles of galaxy clusters and investigate
the influence of the background cosmogony. Models with $\Omega < 1$ and/or
$\Lambda \ne 0$ are considered as well as scenarios with a cold and a hot
component.  Also a pure hot dark matter model is considered as an example of a
cosmogony in which structure does not form hierarchically.  We investigate the
formation history of haloes and its dependence on the background cosmogony. We
also present results for other gross properties such as the triaxiality and
angular momentum of the haloes. Their evolution is compared with the analytical
predictions of the spherical top-hat model.  The structure of the paper is as
follows.  In section 2 we present the different considered cosmogonies and
briefly describe their impact on structure formation. Section 3 presents the
numerical techniques and discusses the issues of numerical resolution, time
stepping and boundary conditions. Section 4 investigates the collapse of
individual haloes in detail showing the evolution of various halo properties.
The structure of the haloes and their dependence on the background cosmogony is
analysed in section 5. We summarise our results and conclude in Section 6.

\section{The cosmological models} \label{lintheo}

Seven cosmological models are studied in this paper. Three of these are cold
dark matter models (CDM, OCDM and LCDM) representing flat, open and $\Lambda$
dominated cosmologies. The three mixed dark matter models (CHDM I-III) contain
differing number and masses of neutrino species.  These models thus cover a
broad range of hierarchical structure formation scenarios.  The HDM model is
added to investigate the formation of clusters in a cosmogony which does not
follow the hierarchical scenario.  Table~\ref{tb1} gives the cosmological
parameters defining the seven different cosmological models.

It is far from obvious what is the best way to compare galaxy clusters as they
form in different cosmological scenarios. Potential choices involve comparing
clusters of similar mass or similar $M/M_*$, among others. The nonlinear mass
$M_*$ is defined as the mass corresponding to the linear rms density fluctuation
which exceeds the critical threshold $\delta_c(z)$ for collapse at a given
redshift in the spherical top-hat model (Eke \etal 1996). Values for $M_*$ at $z
=0$ are given in table~\ref{tb2}.  All these criteria yield a different set of
objects which are compared. The problems are further complicated by some freedom
in the choice of the normalisation of the power spectrum (\eg COBE normalised
versus normalised according to cluster abundance). Finally, high resolution
cluster simulations are still computationally expensive, and a numerical study
is restricted to a few objects per cosmological scenario. Differences in the
formation history due to different realisations of a Gaussian random field can
be larger than those induced by the features of a individual cosmogony. In order
to minimise these sampling effects, we use identical phases for the Gaussian
random field for each of the three considered realisations.  Each realisation
corresponds to one massive cluster at $z=0$.  In the following we will refer to
the realisations as realisation A, B and C.  For the HDM model, 2 simulations
have been performed, corresponding to realisations A and C.

\begin{figure}
\epsfxsize= \hsize\epsffile{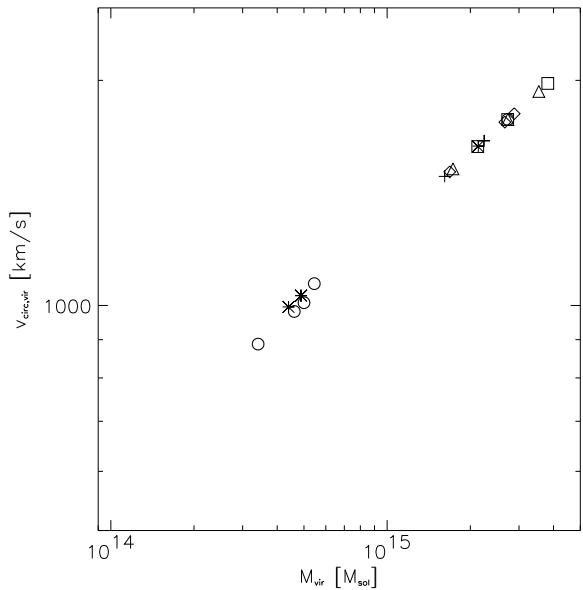}
\caption{\label{models}Mass and circular velocity for the simulated cluster
models.  Circles correspond to clusters formed in OCDM models, stars to LCDM,
plus signs to CDM, triangles to CHDM I, diamonds to CHDM II, squares to CHDM III and crosses to HDM.}
\end{figure}

We normalise all power spectra to $\sigma_8 = 0.63$, i.e. all models have the
same linear extrapolated rms overdensity in top-hat spheres of radius
$8/h\,$Mpc. This normalisation reproduces the observed abundance of rich galaxy
clusters (White \etal 1993) in the case of a high $\Omega$ universe, but is low
for the LCDM and OCDM models.  Recent results of Eke \etal (1996) suggest
$\sigma_8 = 0.5 \; \Omega_0^{-0.52}$ ($\sigma_8 = 0.5
\;
\Omega_0^{-0.46}$) for the LCDM (OCDM). Thus $\sigma_8 = 0.97$ ($\sigma_8 =
0.9$) for $\Omega=0.3$, which is larger than our choice by a factor of $1.5$
($1.4$).  Similar results also hold for COBE normalisation.  While the standard
CDM scenario is inconsistent with the COBE measurements, the CHDM sequences fit
both COBE measurements and cluster abundances. A disadvantage of our
normalisation is that clusters in different cosmological models have different
masses and circular velocitys as shown in figure
\ref{models}. The models with $\Omega < 1$ have approximately a
mass smaller by a factor of $\Omega$ and a circular velocity smaller by
$\Omega^{1/3}$ than the $\Omega=1$ CDM models. Since $M_*$ is smaller by a
similar amount, they have a similar $M/M_*$ as the CDM clusters, \ie they are
similarly rare events at $z=0$.  The clusters of the CHDM models have a much
higher $M/M_*$ and are, therefore, much rarer events.

The simulations are done in a periodic cubic box with a side length of $250 \;
h^{-1}\,$Mpc. The mass resolution is $2.5 \times 10^{11} \; h^{-1} \Omega
\;\Msun$.
Therefore, a cluster with a mass of $10^{15} \; h^{-1} \Omega \; {\mathrm
M_{\odot}}$ consists of 4000 particles. The spatial resolution set by the force
softening and time stepping is $12.5 \; h^{-1}\,$kpc. All simulations are
started at a redshift of $z = 19$ and end at the present time. After every
increase of $0.05$ in the expansion factor an output of position, velocity,
potential and force of the particles is stored, leading to a total of 20 outputs
per simulation.

The initial conditions for the CDM and CHDM runs were generated using the COSMIC
code by Bertschinger \etal (1995). For the LCDM cosmogony the power spectrum
given by Efstathiou
\etal (1992) is used while the OCDM and HDM cosmogonies are based on those
given by Bardeen \etal (1986).

\begin{table}
\caption{\label{tb1} The cosmological parameters of the models.
The columns give the name of the model, the present density parameter of the
cold and hot component of matter, and the density parameter associated with the
cosmological constant $\Omega_\Lambda = \frac{\Lambda}{3 H_0}$, the Hubble
parameter h, and the number of massive neutrinos.}
\centering
\begin{tabular}{lcccccc}
Model & $\Omega_{\mathrm{cold,0}}$ & $\Omega_{\mathrm{hot,0}}$ &
$\Omega_{\Lambda}$  & $h$ & $N_{\nu}$  \\ \hline
CDM & 1.0 & 0 & 0 & 0.5  \\
LCDM & 0.3 & 0 & 0.7 & 0.7  \\
OCDM & 0.3 & 0 & 0 & 0.7   \\
CHDM I & 0.8 & 0.2 & 0 & 0.5  & 3 \\
CHDM II & 0.8 & 0.2 & 0 & 0.5  & 1  \\
CHDM III & 0.7 & 0.3 & 0 & 0.5  & 1  \\
HDM I & 0 & 1.0 & 0 & 0.5 &  1  \\
\end{tabular}
%\end{center}
\end{table}

\begin{table}
\caption{\label{tb2} The effective slope of the power spectrum at
$k = 0.2 \; h \; {\mathrm Mpc}^{-1}$  and the nonlinear Mass $M_*(z=0)$ of the
different power spectra.}
\centering
\begin{tabular}{lcc}
Model           & $n_{\rm eff}$         & $M_{*}/M_{\odot}$
 \\ \hline

CDM             & -1.12         & $1.60 \times 10^{13}$ \\
LCDM            & -1.60         & $4.61 \times 10^{12}$ \\
OCDM            & -1.74         & $5.10 \times 10^{12}$ \\
CHDM I          & -1.64         & $2.26 \times 10^{12}$ \\
CHDM II         & -1.73         & $2.16 \times 10^{12}$ \\
CHDM III        & -2.10         & $1.82 \times 10^{11}$ \\
HDM             & -3.64         & not defined     \\
\end{tabular}
%\end{center}
\end{table}

The standard picture of structure formation involves the gravitational collapse
of density inhomogeneities in an otherwise uniform universe. A simple analytic
description of this process is the spherical top-hat model (Peebles 1980), which
assumes a uniform spherical overdensity. The time evolution of a halo with mass
$M$ and radius $R$ is given by the Tolman-Bondi equation,
\begin{equation}
\dot{R}^2 = 2 G M R^{-1} + \frac{\Lambda}{3} R^2 - 2 E,
\label{TBdi}
\end{equation}
where $E$ is the total energy of the halo and has to be $< 0$ for a collapse.
This model gives the mean over-density $\bar{\delta}(z)$ of a region which is
required for collapse,
\begin{eqnarray}
1 + \bar{\delta}(z) & > &  \frac{1}{\Omega(z)} \qquad \Omega_{\Lambda} = 0 \nonumber \\ 
& > & 1  \qquad  \Omega_0 + \Omega_{\Lambda} = 1.
\label{rhoin}
\end{eqnarray}
In a flat universe any overdense region is able to collapse, whereas in an open
universe $\bar{\delta}(z)$ has to be above a critical value.  The mean density
of an object after virialisation can also be estimated in the spherical collapse
model.  Using the results of Eke \etal (1996) the value of this density,
hereafter virial density, can be adequately approximated for various
cosmological background models by
\begin{equation}
\rho_{\rm vir}(z)  =  180 \; \Omega(z)^{-m} \rho_b(z)\, ,
\label{virial}
\end{equation}
where $\Omega(z)$ and $\rho_b(z)$ are the density parameter and the background
density at redshift $z$. The value of the power index $m$ depends on the
presence of a cosmological constant. For $\Lambda = 0$ $m = 0.66$, while for
$\Omega+\Omega_\Lambda =1$, $m =0.52$.

The spherical collapse model is, however, a strong simplification for structure
formation in hierarchically clustering scenarios. Tracing back collapsed
structures to high redshift shows that structures form from highly irregularly
shaped volumes (see \eg figure
\ref{figala}) which also possess a large amount of substructure. The
formation of a halo of the size of a cluster must not be only the result of
spherical accretion of matter shells but can also be formed by merging of
smaller matter accumulations. Generally the evolution pattern is influenced by
the overall distribution of the inhomogeneities and the cosmological background
model. In the following we will discuss how the cosmological background model
and the kind of dark matter affect the formation of a halo.

There is an important difference between the models based on an
Einstein-de~Sitter universe and the models with $\Omega < 1$ concerning the time
evolution of the density inhomogeneities.  At high redshifts when linear theory
is valid the growth of overdensities can be described by $\delta({\bf x},z) =
D(z)
\delta_0({\bf x})$ (Peebles 1980). $\delta({\bf x},z)$ is the local density
contrast at redshift $z$, $\delta_0({\bf x})$ the linearly extrapolated value at
the present time and $D(z)$ the linear growth factor.  It gives the growth rate
of the overdensities in linear theory.  
In an EdS universe $D(z)=a=1/(1+z)$.  In a universe with $\Omega < 1$, the slope
of $D(z)$ is shallower for low $z$. Since the growth factor is normalised to
$D(0) = 1$, $D(z)$ is thus larger at higher redshifts and the nonlinear regime
is reached earlier than in the models based on the EdS universe. For $\Omega =
0.3$, $D/a \approx 2$ for an open universe ($\Lambda = 0$) at a redshift of $z
=19$, and $D/a \approx 1.3$ for a flat universe ($\Lambda = 1 -
\Omega$).  However in an open universe an overdense region needs a higher
density contrast to collapse than in a flat universe due to
equation~(\ref{rhoin}). This shifts the beginning of structure formation to
smaller $z$-values in an open universe, partially neutralizing the effect of a
higher growth factor relative to a flat universe with the same density
parameter.

\begin{figure}
\centering
\epsfxsize= \hsize\epsffile{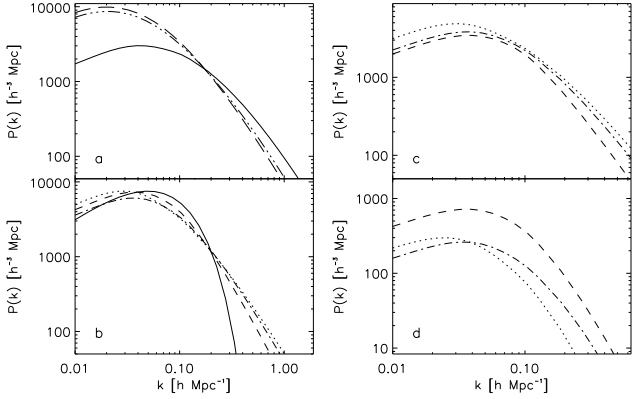}
\caption{Power spectra $P(k)$ of all cosmoginic models are compared at $z =19$:
a) $P(k)$ of the CDM (solid), LCDM, (dash triple-dotted) and the OCDM model
(long-dashed); b) the total power spectra of the three CHDM models (I: dotted,
II: dash dotted, III: dashed) and the HDM model (solid); c) and d) $P(k)$ for the two components of the mixed dark matter models seperatly, c) for the cold and d) for the hot component.}
\label{figps}
%\end{center}
\end{figure}

The general distribution of inhomogeneities depends on the shape of the power
spectrum $P(k)$, which in turn is set by the density parameter and the physical
properties of the assumed dark matter. In figure~\ref{figps} the power spectra
of all cosmogonies used are plotted. On the basis of the asymptotic behavior of
the power spectra at high wave numbers one can distinguish between two different
structure formation scenario: the hierarchical scenario and the non-hierarchical
(top-down) scenario.

Power spectra with an asymptotic slope larger than $-3$ have more power on small
scales. This feature is believed to lead to a hierarchical scenario, where small
objects collapse first and large object like clusters form later by merging of
smaller ones.  All models except for the HDM cosmogony follow this pattern.  The
detailed evolution pattern in the hierarchical scenario depends on the effective
slope $n_{\rm eff}$ of the power spectrum near $k_{\rm cl}$, the wave vector
which corresponds to the mass of a considered cluster.  For a cluster mass of
$10^{15} \; h^{-1} \Omega \; \Msun$ and $\sigma_8 = 0.63$ this is $k_{\rm cl}
\approx 0.2 \; h \;$Mpc$^{-1}$. Since $k_{\rm cl}$ is similar to the wave number
at which the power spectra have been normalised, the amplitude of $P(k)$ is
similar at $k_{\rm cl}$. The degree of inhomogeneity on small scales is related
to the value of $n_{\rm eff}$.  Values of $n_{\rm eff}$ are also given in
Table~\ref{tb2} for all models. For similar overdensities under equal
cosmological conditions small scale structure is more pronounced in cosmogonies
with larger values of $n_{\rm eff}$ or equivalently of $M_*$.  Small scale
objects are more numerous and form at earlier redshift, leading to a higher
merging rate of these objects than in models with small $n_{\rm eff}$ or
$M_*$. Within our models the most small scale power is in the CDM and the
smallest in the CHDM III simulations. The steeper slope of the OCDM and LCDM
models compared to SCDM are almost compensated by the initially faster growth of
the fluctuations.

In the mixed dark matter simulations gravitational clustering is more complex
due to the mutual interaction between the hot and cold components. Due to free
streaming, the hot component is initially much more homogeneously distributed
than the cold component. The smaller overall overdensities also suppress the
growth of density fluctuations in the cold component which thus proceeds slower
than $\propto (1+z)^{-1}$, the linear growth rate for a pure CDM model. Only at
low redshifts has the velocity dispersion of massive neutrinos decreased
sufficiently due to the expansion of the universe that the neutrinos are able to
fall into and virialise within the potential wells formed by the cold dark
matter. The details of the power spectrum and the non-linear clustering thus
depend on the mass fraction of the hot dark matter as well as on the mass of the
neutrinos themselves, \ie on the number of families of massive neutrinos (for a
given mass fraction of hot dark matter).  This dependence can be seen in the
power spectra of the cold and the hot component (Figure~\ref{figps} c,d). The
suppression of large scale power in the cold component is most pronounced in the
CHDM III model due to the high mass fraction in neutrinos ($\Omega_{\rm hot} =
0.3$). Suppression of small scale structure is most pronounced for the CHDM I
model. It has three families of massive neutrinos and, therefore, the lowest
neutrino mass ($m_\nu = 6 h^2\,eV$).

In the case of the HDM model the asymptotic slope is less than $-3$. One cannot
specify $M_*$ since the linear rms density fluctuation for HDM (normalised to
$\sigma_8 = 0.63$) never exceeds the critical threshold for spherical collapse.
Some peaks however are sufficiently high and exceed $\delta_c(z)$.  Large scale
modes become nonlinear first in HDM simulation, leading to the formation of
clusters before galaxy sized objects are present.  This should cause the
evolution of haloes to be qualitatively very different in the HDM
cosmogony. Since there is no small scale power in the initial spectrum, the
formation of a halo should be similar to the collapse of an overdense region in
the spherical top-hat model.  However, this qualitative expectation needs to be
checked by examining the evolution of HDM clusters in the simulations.

\section{Numerical techniques}

It is computationally challenging to perform numerical simulations which allows
one to analyse the density distribution of galaxy clusters in sufficient detail.
Such simulations must not have a spatial resolution worse than a couple of tens
of kpc.  In order to include the tidal field exerted by a cosmologically
representative sample of the universe, however, simultaneously a box with a side
length of several hundreds of Mpc has to be covered.  The high spatial
resolution also requires a fine grained time stepping which gives rise to a
total number of time steps of more than $10^4$.  In the traditional approach of
large scale N-body simulations using (adaptive) P3M (see,
\eg Efstathiou \etal 1985, Couchman 1991), such simulations, if at all
possible, require the computing power of state of the art massively parallel
supercomputers.

As an alternative to large N-body simulations, multi-mass techniques have been
developed (Porter 1985).  In analogy to the concept of tree codes, the tidal
field of the surrounding matter is represented by particles whose mass increases
logarithmically with distance.  These techniques have been successfully applied
to the formation of galaxies and of galaxy clusters (Katz \& White 1993,
Navarro, Frenk \& White 1995, Bartelmann, Steinmetz \& Weiss 1995).  Its
disadvantage is that only one or a few objects can be studied per simulation,
and the selection of the region which potentially forms an object involves some
bias.

In this section we describe a new variant of such a multi-mass technique, using
nested distributions of particles with different mass. Additional a high
resolution N-body integrator is combined with a low resolution particle-mesh
(PM) code.  
Such a hybrid technique combines the advantages of PM and high resolution codes:
(i) it naturally provides periodic boundary conditions, (ii) the forces within
the high resolution follow a $r^{-2}$ law, smoothed by a plummer or a spline
softening and (iii) a smooth sampling of the external force field surrounding
the object of interest is given.  Two-body relaxation due to accidental
encounters of particles which largely differing mass can be suppressed, which
may be of even more importance for simulations including the effects of gas
dynamics (Steinmetz \& White 1996).  The periodic extension under preservation
of the Newtonian force law is especially interesting for codes which use the
special purpose hardware GRAPE (Sugimoto \etal 1989), where a plummer force law
is hardwired.  But also in the case of tree codes such a feature is of interest.
In contrast to the Ewald summation technique (Ewald 1921, Hernquist \etal 1991),
it easily and memory efficiently allows one to combine periodic boundary
conditions and higher order multipole moments calculating the tree force.

\subsection{Initial conditions}

\begin{figure}
\centering
\epsfxsize= \hsize\epsffile{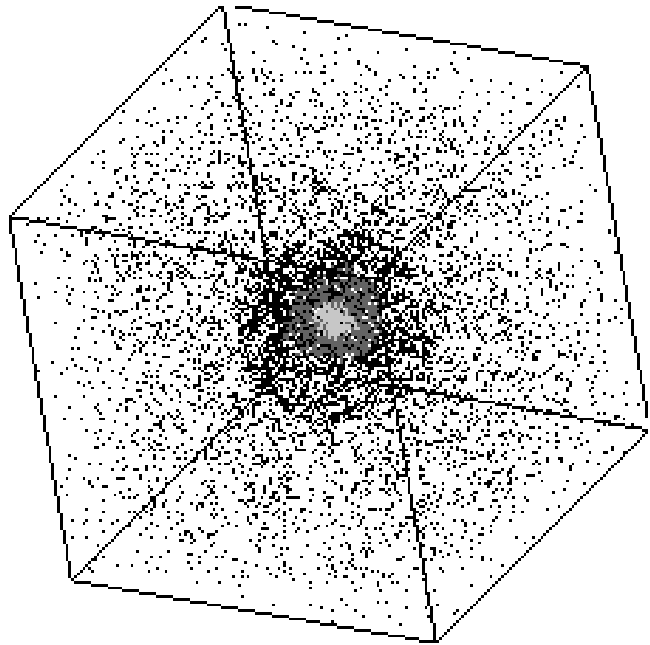}
\caption{Initial particle configuration showing the hierarchy of four
nested
particle distributions.}
\label{figdp}
\end{figure}

The initial particle arrangement consists of a hierarchy of four nested particle
distributions (see figure~\ref{figdp}, \ref{figar}). The particle mass increases
by a factor of eight going from one hierarchy to the next. The hierarchy with
the lowest resolution covers the simulation box with periodic boundary
conditions and a side length of $250 \; h^{-1} \;$Mpc.  A sphere with a radius
of $1/16$th of the box size is the highest resolution region. All particles
which can be found within the virial radii of galaxy clusters studied in this
paper are acquired from this sphere. The highest and lowest resolution regions
are connected by two intermediate spherical shells of particles, the outer
radius of these shells being $1/8$th and $1/4$th of the box size, respectively.
In the central region the dynamic range of the simulation therefore corresponds
to a simulation with $256^3$ particles per dark matter species, however, the
actual particle number is only $\approx 3
\times 64^3$.

The initial particle distribution within each hierarchy consists of a
glass-like particle arrangement, i.e.~the force on every individual
particle equals zero. 
Like a Poissonian sample, and in contrast to a grid, such a particle arrangement
has no preferred direction, but contrast to a Poissonian sample, it also has no
considerable large scale power up to the Nyquist frequency (Efstathiou \etal
1995).  It can easily be created assuming repulsive rather than attractive
interparticle forces (White 1996). Each level uses the same glass-like particle
arrangement but with different mass assigned to the particles.

In the case of models with hot and cold dark matter, each component is
represented by a separate particle distribution. The mass of each particle is
scaled according to the relative contribution of each component. In order to
prevent particles having nearly identical initial positions the distributions
for each species is drawn from a different realisation of a glass-like particle
arrangement.

To realise a Gaussian random field with a given power spectrum, the position of
every particle is perturbed according to the Zel'dovich approximation
(Zel'dovich, 1970).  For the lower mass hierarchies, the displacement vectors
consist of the identical wave numbers as used for the surrounding distributions
and of additional wave numbers up to the Nyquist frequency of the considered
hierarchy. In the case of a hot dark matter component, a randomly oriented
thermal velocity is added (Klypin \etal 1993).

We also use Zel'dovich approximation in order to estimate the region within
which a galaxy cluster is potentially forming.  Based on a low resolution run
(mass of a particle equal to the particle mass of the highest mass hierarchy)
particles are propagated to redshifts $z \approx 0.5 - 0$.  The final redshift
is chosen such that structures are not yet washed out do to shell crossing.  The
simulation box is recentered to the potential formation place of a galaxy
cluster.  Afterwards, particles within the high resolution region are
successively replaced by the lower mass hierarchies, new wavelength are added,
new displacements are calculated and the box is recentered, accordingly.  This
method turns out to be a reliable and computational inexpensive way to identify
regions within which (rich) galaxy clusters are forming.

\subsection{The N-body code}

\subsubsection{The hybrid PM/high-resolution code}

Our N--body combines a low resolution PM code with high resolution scheme (\eg a
tree code or a PP code using the special purpose hardware GRAPE).  This hybrid
scheme enables us to use a large box size and periodic boundary conditions but
simultaneously also to incorporate accurate short range forces within the high
resolution region (throughout this section we label the mutual forces between
particles of the lower three mass hierarchies as short range forces).

\begin{figure}
\centering
\epsfxsize=\hsize\epsffile{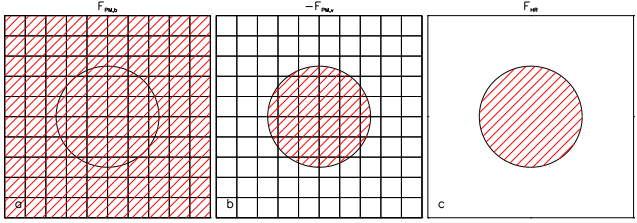}
\caption{Sequence of the force calculations for the hybrid
PM/high resolution (HR) code.  The inner sphere corresponds to the HR region,
the shaded area indicates those parts of the simulation box which are involved
in the force calculation; a) The PM force with periodic boundary conditions
${\bf F}_{\rm PM,p}$ is computed for all particles, b) only for particles of the
lower three mass hierarchies, the PM force with vacuum boundary conditions ${\bf
F}_{\rm PM,v}$ is calculated and subtracted from ${\bf F}_{\rm PM,p}$, c) the HR
code is used to calculate high resolution forces ${\bf F}_{\rm HR}$ again only
for the particles of the lower mass hierarchies.  The total force for these
particles is given by ${\bf F}_{\rm PM,p}-{\bf F}_{\rm PM,v} + {\bf F}_{\rm
HR}$.}
\label{figfc}
\end{figure}

The force calculation involves three computational steps (see
figure~\ref{figfc}): In a first step the PM forces ${\bf F}_{\rm PM,p}$ are
calculated for all particles within the simulation box assuming periodic
boundary conditions.  In a second step, the low resolution PM forces exerted by
particles of the lower three mass hierarchies ${\bf F}_{\rm PM,v}$ are
calculated assuming vacuum boundary conditions.  This force is subtracted from
${\bf F}_{\rm PM,p}$ for all particles of the lower three mass hierarchies.  The
difference $\Delta{\bf F}_{\rm PM} = {\bf F}_{\rm PM,p} - {\bf F}_{\rm PM,v}$
gives the periodic extension of the force for these particles.  In a third step,
the short force ${\bf F}_{\rm HR}$ is calculated by means of a high resolution
scheme using vacuum boundaries.  In summary, the hybrid scheme replaces the PM
short range forces by a PP or a tree code, but keeps the low resolution PM
forces for all periodic images.

The PM forces ${\bf F}_{\rm PM,p}$ and ${\bf F}_{\rm PM,v}$ are calculated on
$32^3$ mesh via Fast Fourier transforms. Particles are assigned to the grid
using TSC assignment (Efstathiou \etal 1985). Since for ${\bf F}_{\rm PM,v}$
less than 1/8th of the volume of the mesh is occupied, vacuum boundaries can be
easily implemented in the PM calculation (Eastwood \& Brownrigg 1979).  Beside
the periodic images, the mesh forces within the high resolution region is
exactly zero and the short range forces obey a Newtonian force law.  Therefore,
our hybrid scheme is different from the classical P3M approach, where the short
range force is a correction to the mesh force and, therefore, it does not follow
a $1/r^2$ law. Furthermore, no matching error at the maximum radius of the PP
correction is involved (Efstathiou \etal 1985).

\begin{figure}
\centering
\epsfxsize= \hsize\epsffile{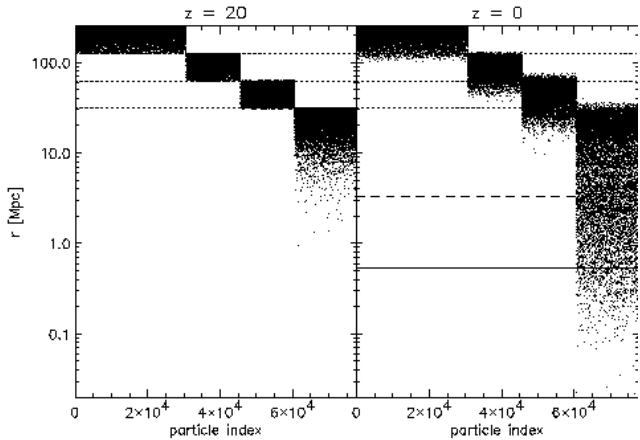}
\caption{Radius $r$ of a particles at the begin (left) and the end of the
simulation.  The dashed line corresponds to the virial radius, the solid line to
the scale radius of the cluster at $z=0$. The three dotted lines correspond to
the outer radii of the lower three mass hierarchies at $z=z_{\rm init}$. As can
bee seen, mixing of particles across the boundary of the highest mass hierarchy
is only weak.  No particle of the upper three mass hierarchies can be found
within $3 r_{\rm vir}$.}
\label{figar}
\end{figure}

Throughout a simulation, the way how forces are calculated are not changed for
every individual particle: for the highest mass hierarchy, the force is always
${\bf F}_{\rm PM,p}$, for the lower three mass hierarchies always ${\bf F}_{\rm
PM,p}-{\bf F}_{\rm PM,v} + {\bf F}_{\rm HR}$.  No discontinuities matching two
different numerical schemes together are involved.  A potentially pathological
case may occur, if particles of the lower mass hierarchies move outside the
central box of side length $r_{\rm box}/2$.  The vacuum PM force for those
particles is then not properly represented.  An additional tidal field arises
since the periodic images of these particles are not properly compensated.  The
high resolution region, however, is situated around a galaxy cluster and it
therefore covers an overdense region.  Only rarely, if at all do particles of
the lower three mass hierarchies leave the central sphere of radius $r_{\rm box}/4 =
62.5\,h^{-1}$Mpc, as also can be seen in figure~\ref{figar}.  Particles of the
highest mass hierarchy which move into the high resolution sphere are less
problematic.  Due to the low resolution of the PM force, their perturbing effect
is rather limited.  As can be seen in figure~\ref{figar}, the mixing between
different mass hierarchies is rather limited and only neighbouring hierarchies
interact.  The largest mass difference between closely interacting particles is
thus only factor of eight and relaxation effects are only weak.  Furthermore it
can be seen that the minimum radius which a particle of the second mass
hierarchy achieves is still a factor of 3 larger than the virial radius of the
galaxy cluster at $z=0$.  Thus the galaxy cluster and its environment is
consistently simulated with the highest resolution and perturbing relaxation
effects due to accidently intruding more massive particles can be excluded.

Finally we note, that the level within the mass hierarchy at which the force
calculation is switched from the PM to the high resolution force is to some
extent arbitrary. For the combination of PM with a GRAPE-PP code (this code is
used for the simulations presented in this paper) it is computationally
advantageous to use PM only for the highest mass hierarchy. The PM part has
thoroughly to be performed on the (relatively slow) front end, while the PP
part benefits from the high performance of GRAPE. For a similar code on a
conventional computer, which uses a tree code for the short range forces, the
opposite is true: the calculation of the short range forces is more expensive
than the PM part. It is likely to be more advantageous to use a finer grid and
to calculated not only the highest mass hierarchy by PM but also some of the
lower ones.

\subsubsection{The influence of gravitational softening, time stepping and
boundary conditions}
The high resolution force is softened according to a Plummer law.  The softening
length is set to $\epsilon_{\rm ref} = 12.5\; h^{-1}\,$kpc.  The time evolution
of the N-body system is integrated by a leap frog time integrator.  The time
step is set to
\begin{equation}
\Delta t_{\rm ref} = \frac{\epsilon_{\rm ref}}{\max_i (|{\bf v}_i|)}\, ,
\end{equation}
${\bf v}_i$ being the velocity vector of particle $i$.  Because the PM
calculation consumes a substantial part of the total CPU time per time step, we
do not use an individual time step scheme.

In order to study the influence of numerical parameters, smoothing length and
time step criteria are varied and their influence on the structure of a halo is
investigated.  For CPU economy reasons, these test simulations use only
particles of the lowest mass hierarchy and vacuum boundary conditions.  Initial
displacements and velocities are identical to the CDM C realisation.

\begin{figure*}
\centering
\hskip0.1\hsize\epsfxsize=0.8\hsize\epsffile{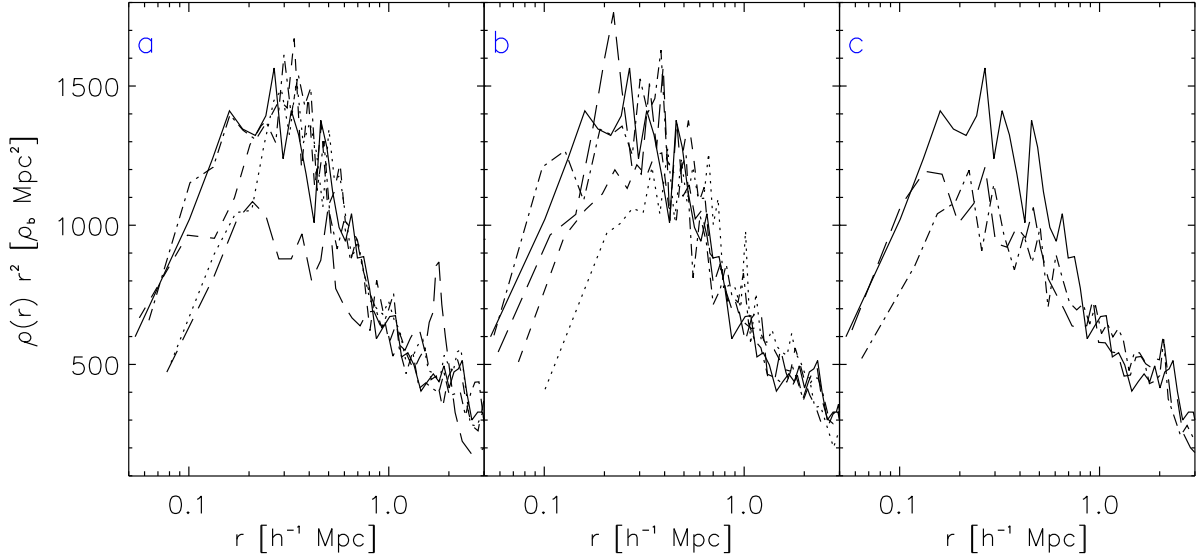}
\vskip1cm
\caption{The influence of softening, time stepping and boundary conditions are
tested on the density stratification inside a halo. The profiles of
$\varrho\,r^2$ as function of $r$ are plotted for different test
simulations. Plot a) shows the profiles for different plummer softening,
$\epsilon = \epsilon_{\rm ref}$ (solid), $2 \epsilon_{\rm ref}$ (dashed), $0.4
\epsilon_{\rm ref}$ (dash dotted), $8
\epsilon_{\rm ref}$ (dotted) and $0.2 \epsilon_{\rm ref}$ (long-dashed
dotted), plot b) for different time stepping: $\Delta t = \Delta t_{\rm ref}$
(solid), $0.25 \Delta t_{\rm ref}$ (dash dotted), $4 \Delta t_{\rm ref}$
(long dashed), $8 \Delta t_{\rm ref}$ (dashed) and $10 \Delta t_{\rm ref}$
(dotted), plot c) for different boundaries conditions: isolated boundaries at
$r=r_{\rm box}/16$ (solid) and at $r=r_{\rm box}/4$ (dashed) and periodic
boundaries (dash dotted).}
\label{figsmd}
\end{figure*}

In figure \ref{figsmd}a we demonstrate the effect of gravitational softening on
the density stratification inside a halo plotting the profiles of $r^2
\varrho$. The binning of the particles is described in detail in
section~3.3.2.  The softening length $\epsilon$ is 
varied between $0.2
\epsilon_{\rm ref}$ and $8 \epsilon_{\rm ref}$.
To fix the number of time steps to nearly the same value (about 4000 time steps)
the time step is set to $\Delta t = \beta \frac{\epsilon}{\max_i (|{\bf v}_i|)}$
with $\beta$ varying between 5 and 0.125 correspondingly.  
The reference softening $\epsilon_{\rm ref}$ corresponds to less than 1\% of the
virial radius of the test halo.  As can be seen, the results are fairly
converged for a softening between $0.4\epsilon_{\rm ref}$ and $\epsilon_{\rm
ref}$.  For the model with $\epsilon = 2\epsilon_{\rm ref}$ slight while for the
model with $\epsilon = 8\epsilon_{\rm ref}$ a substantial depletion of the
density near the center ($r < 0.3\, h^{-1}$ Mpc) can be observed, though the
density distribution at large scales is still fairly well represented.  In the
case of the smallest softening length $\epsilon = 0.2\epsilon_{\rm ref}$
significant deviations can be observed.  This demonstrates the importance of the
number of time steps, the simulation has too small a number of time steps to
handle the strong interparticle forces which can arise for such a small
softening length.  Close two body encounters result in high energy particles
which escape the halo.

In figure \ref{figsmd}b the influence of the time stepping is investigated.
Fixing the smoothing length to $\epsilon = \epsilon_{\rm ref}$ five simulation
are done with $\Delta t = \Delta t_{\rm ref}$, $\Delta t = 0.25 \Delta t_{\rm
ref}$ $\Delta t = 4 \Delta t_{\rm ref}$ $\Delta t = 8 \Delta t_{\rm ref}$ and
$\Delta t = 10 \Delta t_{\rm ref}$.  The profiles of $\varrho\,r^2$ for $\Delta
t \le 4 \Delta t_{\rm ref}$ show no clear deviation from each other.  But for
$\Delta t > 4 \Delta t_{\rm ref}$ the central region is not as compact as for
the smaller time step criteria. This indicates that the chosen time stepping is
sufficiently finely grained.

The influence of the boundary conditions is investigated by comparing three
simulation (initial conditions as in model CDM C) under differing boundary
conditions.  One simulation uses periodic boundary conditions (BC), one uses
vacuum boundaries at the boundary of the highest resolution sphere (inner
$31.25\,h^{-1}$Mpc sphere) (VC1) and one vacuum boundaries at the boundary of
the third mass hierarchy (inner $125\,h^{-1}$Mpc sphere) (VC3). The simulation
BC was performed with the hybrid PM/GRAPE scheme presented above, while for the
runs VC1 and VC3 the PM part has been disabled. All other numerical parameters
are kept identical for all runs, \ie $\epsilon=\epsilon_{\rm ref}$ and $\Delta
t=\Delta t_{\rm ref}$.

In figure~\ref{figsmd}c, $\varrho\,r^2$ is compared for all three models. They
agree well above $r \approx 1\, h^{-1}$ Mpc but show clearly differences in the
central region. The halo center becomes less dense with increasing surrounding
matter and even a slight difference in the central density between model BC and
VC3 can be observed.  This demonstrates the importance of properly including
large scale matter distribution in such simulations.  As we will discuss in
section 5.3, the gravitational field exerted by surrounding matter may play a
critical role for the density distribution of dark matter haloes.

In summary these results show that the influence of different numerical
parameters is fairly well understood.  For parameters close to the reference
values proposed above, the density profiles are consistent with each other.
Within the numerical discretisation errors, the choice of a specific code and/or
a specific type of softening does not affect the results of the simulations.
Nevertheless, the total number of time steps of about $10^4$ is still moderate
and allows us to investigate a fairly large sample of cosmological models, each
of which represented by several sets of initial conditions.

\subsection{Miscellaneous}

\subsubsection{Identification of haloes and substructure} \label{IHS}

Later in this paper, the formation history of haloes will be investigated.  The
mass of the most massive progenitor as well as the mass in collapsed structure
will turn out to be a valuable tool to characterise the merging histories of
individual haloes.  In order to identify collapsed substructures and to label
those particles which belong to such a structure, we use a group finding
algorithm which is based on a binary tree structure similar to that of a N-body
code.  In a binary tree, pairs of mutually nearest neighbours are grouped to
nodes, mutually neighbouring nodes to more massive nodes and so on.  These
procedure is iteratively repeated until only one node containing all particles
remains (for details see Porter 1985; Jeringhan \& Porter 1987; Steinmetz \&
M\"uller 1993).  The nodes preferentially cover regions of high particle
density. The estabilshed hierarchy of nodes is now scanned for nodes which can be
identified as bound objects.  Nodes which do not have at least virial density
$\varrho_{\rm vir}(z)$ or which consist of less than $10$ particles are
eliminated. The volume of a node is approximated by the volume of its ellipsoid
of inertia.  From the remaining list of nodes, all those are discarded which do
not enclose a compact mass accumulation. The compactness of a node is inferred
by measuring the mass within $r_{\rm core} = 25\; h^{-1}\,$kpc near the center of
mass of the considered node. The average density within $r_{\rm core}$ is then
required to be larger than the average density of the considered node.
Furthermore any subnode of these nodes must not be underdense.  This procedure
delivers a list of nodes (i) which are not a subset of a compact, more massive
node and (ii) which are part of collapsed bound structures. The advantage of our
new algorithm is that close halo pairs can be resolved.

\subsubsection{Radial matter distribution of haloes}

In section~\ref{CP0} we will discuss the matter distribution of haloes.  The
density profile is measured by binning the particle distribution in 200
spherical shells centered on the minimum of the cluster potential.  Each shell
contains $N$ particles, where $N$ is logarithmically increasing with radius
starting with $N=120$ near the center. At $z=0$, the bin at $r_{\rm vir}$
contains about $140$ particles.
%%%### stimmen die 140 ?
For the mixed dark matter models twice as many particles per bin have been
used. Each bin is assigned a radius corresponding to the average distance of all
particles which belong to the bin. This radius is smaller than the half-volume
radius of a shell, since the density decreases with distance from the cluster
center.  The use of spherical bins for triaxial structures like clusters has no
affect on the shape of profiles (Lemson 1995).

The radial density distribution can sensitively depend on the definition of the
halo center.  We use the minimum of the gravitational potential to define the
halo center. This halo center usually 
coincides with that defined by the maximum density or the center of mass, but
for bimodal mass distributions (as typically occur during merging events) they
can differ substantially. While for a bimodal mass distribution the minimum of
the gravitational potential focuses on one of the two subclumps, the center of
mass is located between the two contributing subclumps. Binning the mass
distribution in radial shells centered on the center of mass thus results in an
artificially shallow density distribution near the halo center. Compared to the
maximum density, the minimum of the potential has the advantage to select the
most massive progenitor as the center, while the maximum density can correspond
to a small clump of high central density.

The potential minimum of cluster is defined in the following way: The
gravitational potential of each contributing particle is assigned to a $32^3$
cube grid using NGP charge assignment (Eastwood \etal 1980) and the mesh cell
with the lowest potential is identified. The center of a halo is then defined as
the mean value of the potential weighted positions of all particles inside a
sphere around the cell center with a radius equals to half the cell size.

The outer boundary of a cluster is defined by the virial radius $r_{\rm
vir}$, the radius at which the mean overdensity within a sphere equals the
virial density $\varrho_{\rm vir}$ of the spherical top-hat
model (equation 5). All particles which lie inside $r_{\rm vir}$ are taken to belong to the
cluster.

\section{The process of structure formation} \label{PSF}

This section investigates merging history, shape and angular momentum of
clusters. Variations between different clusters of a given scenario, as well as
between clusters from different cosmogonies are discussed. We consider first the
qualitative nature of cluster collapse as it arises in our simulations.

\subsection{Mergers and mass accretion}\label{mma}

\begin{figure*}
\centering
\hskip0.05\hsize\epsfxsize=0.9\hsize\epsffile{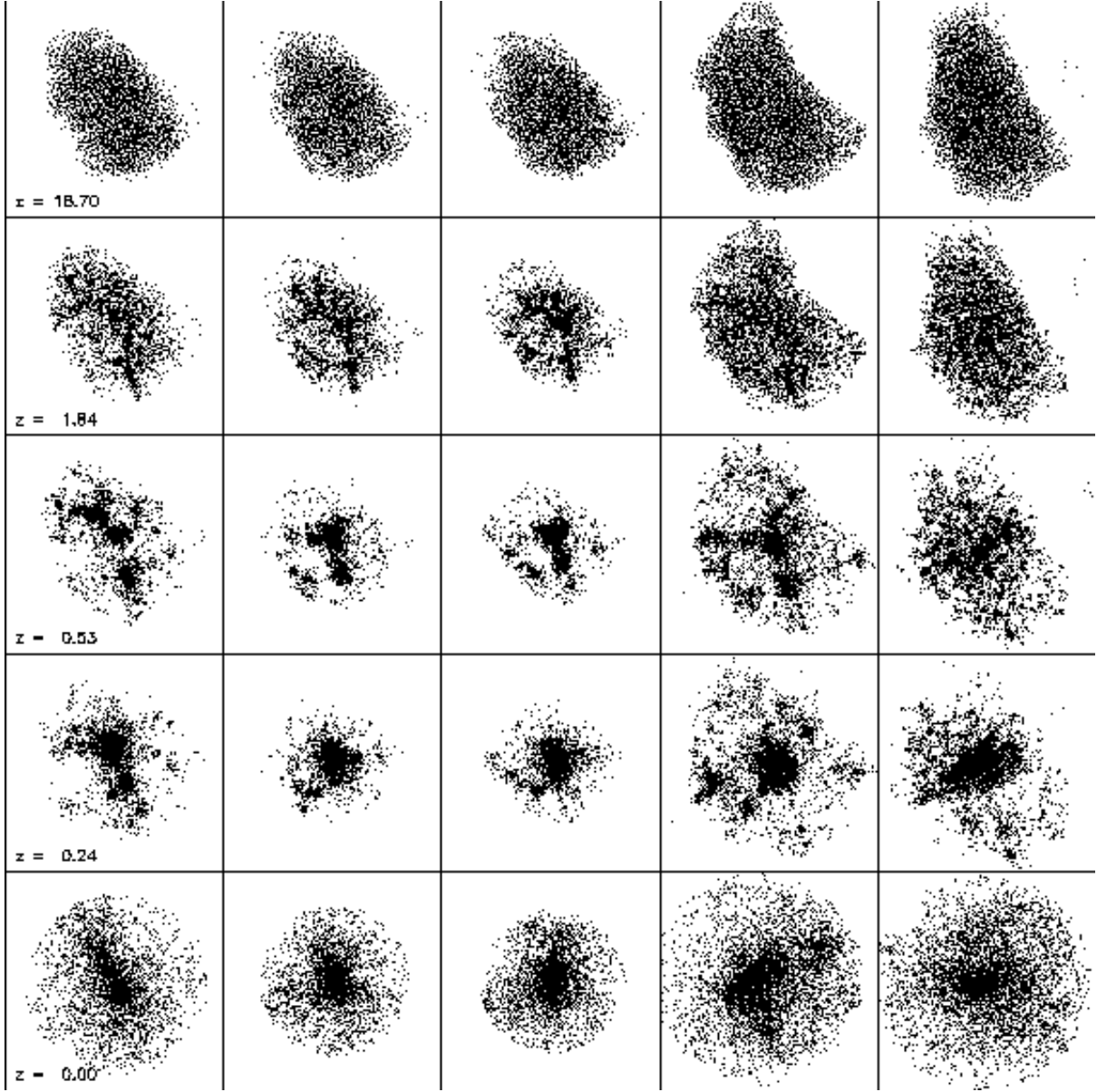}
\caption{Distribution of particles of cluster A (at $z =
0$ ) of the CDM, LCDM, OCDM, CHDM III and HDM 
cosmogony at $z = 19$, $1.8$,
$0.5$, $0.2$ and $0$ (from above). The physical size of the corresponding boxes
is $1.7$, $9.1$, $11.4$, $9.5$ and $4.5 \, h^{-1} {\mathrm Mpc}$.}
\label{figala}
%\end{center}
\end{figure*}

The formation of a cluster can be followed by looking at the distribution of
those particles which at the present epoch lie within a sphere of radius $r_{\rm
vir}$.  In figure~\ref{figala}, the particle positions projected in the x-y
plane are shown for the realisation A of the CDM, LCDM, OCDM, CHDM III and HDM
models. The particle distribution is shown at redhsifts $z=18.7, 1.84,
0.53, 0.24, 0$.  The selected models represent typical evolution patterns for
the various cosmological parameters.  Some differences between the CDM, LCDM and
OCDM can be seen, in particular the earlier collapse and greater compactness of
the OCDM cluster at $z=1.84$ and $0.53$. The largest differences are between
various cold dark matter clusters on the one hand, and the CHDM III and HDM
clusters on the other. The latter models form distinct structures only at
redshifts $z<1$.

%\clearpage

In the cold dark matter models the clusters of each realisation follow coarsely
the same evolution pattern.  Covering initially an ellipsoidal shaped region the
particles in realisation A accumulate first along filamentary structures. Within
the filaments several small spherical shaped objects of similar mass are
formed. These structures preferably merge at the intersections of the
filaments. A few clumps of successively larger mass are built up, and following
a central impact they merge together. Smaller clumps are tidally disrupted
during the infall on the central matter accumulation.  The nonlinear evolution
is fastest in the OCDM model, where distinct objects are visible in the
filaments already at $z \approx 4$.  Roughly the same degree of clustering is
reached in all EdS-models at later times. This holds even for the mixed dark
matter models, where the presence of a hot component delays the growth of
structure. The similarity of the clustering pattern at late epochs is largely
due to the normalisation of all models to $\sigma_8=0.63$, \ie the linear
overdensity is identical on a length scale only moderately larger than a typical
cluster scale.

\begin{figure*}
\centering
\hskip0.05\hsize\epsfxsize=0.9\hsize\epsffile{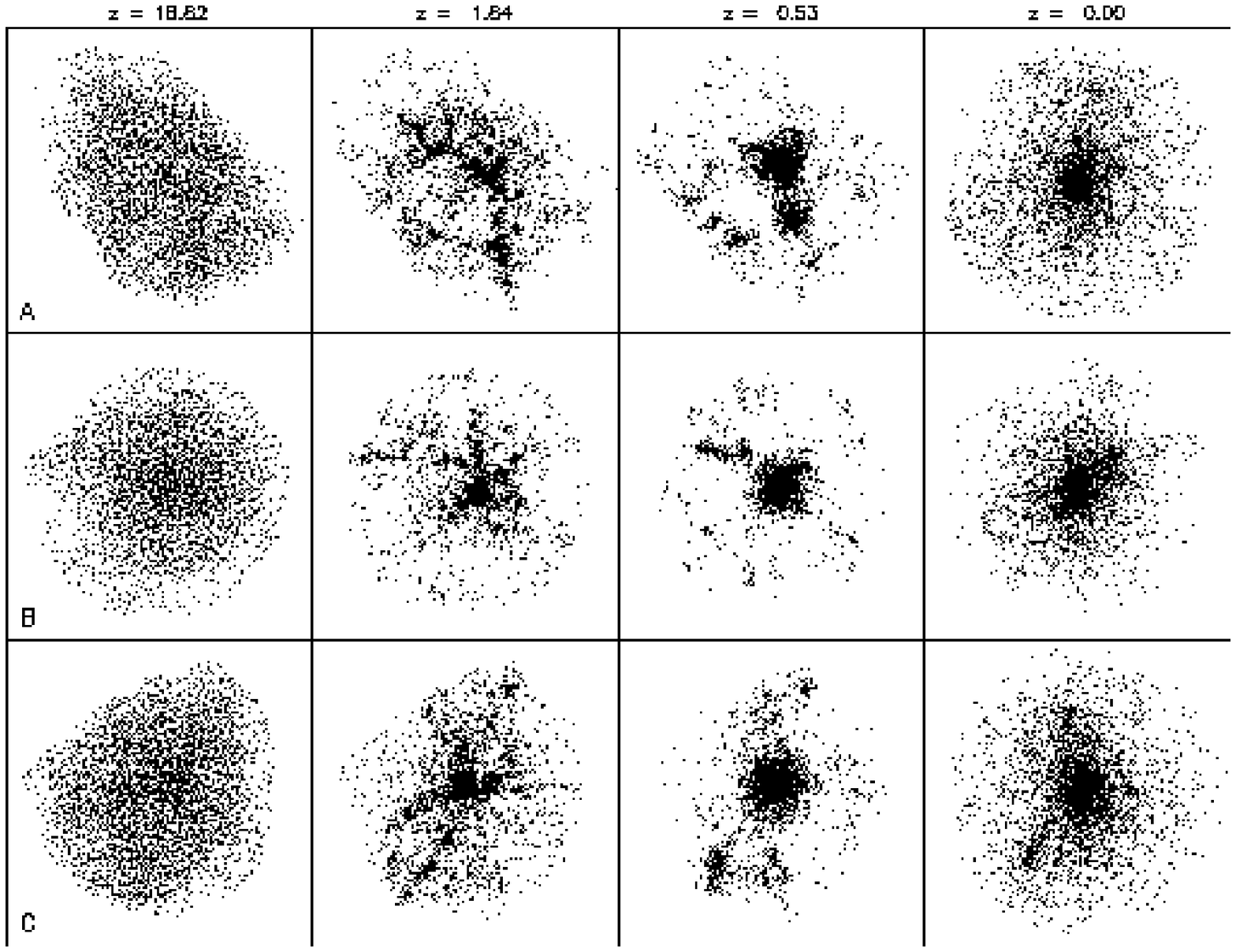}
\caption{Distribution of particles of the most massive cluster (at $z =
0$ ) of all three OCDM realisation A (upper row), B (middle row) and C
(lower
row) at $z = 19$, $1.8$, $0.5$, and $0$ (from left). The physical size of the
corresponding boxes is $1.4$, $6.4$, $7.7$ and $2.9 \, h^{-1} {\mathrm Mpc}$.}
\label{figalb}
%\end{center}
\end{figure*}

However, the formation history can be quite different for different
realisations.  This is shown in figure~\ref{figalb}, where the formation history
of the OCDM cluster is shown for all three realisations A, B and C. The
initial particle distribution seem to be more spherical for realisation B and C,
although this is partly a projection effect. Realisation B is elongated in the
y-z plane by an amount similar to the x-y elongation of realisation A, while
realisation C is less elongated in both projections. At lower redshifts, the
particle distribution also appears less filamentary for realisations B and C,
but for realisation B the orientation of the filament is along the line of
sight. In realisation B and C, a massive clump has established relatively early
near the center and the future evolution of the cluster can be well
characterised by accretion of outer matter shells, similar to the spherical
collapse model.

\begin{figure*}
\centering
\hskip0.05\hsize\epsfxsize=0.9\hsize\epsffile{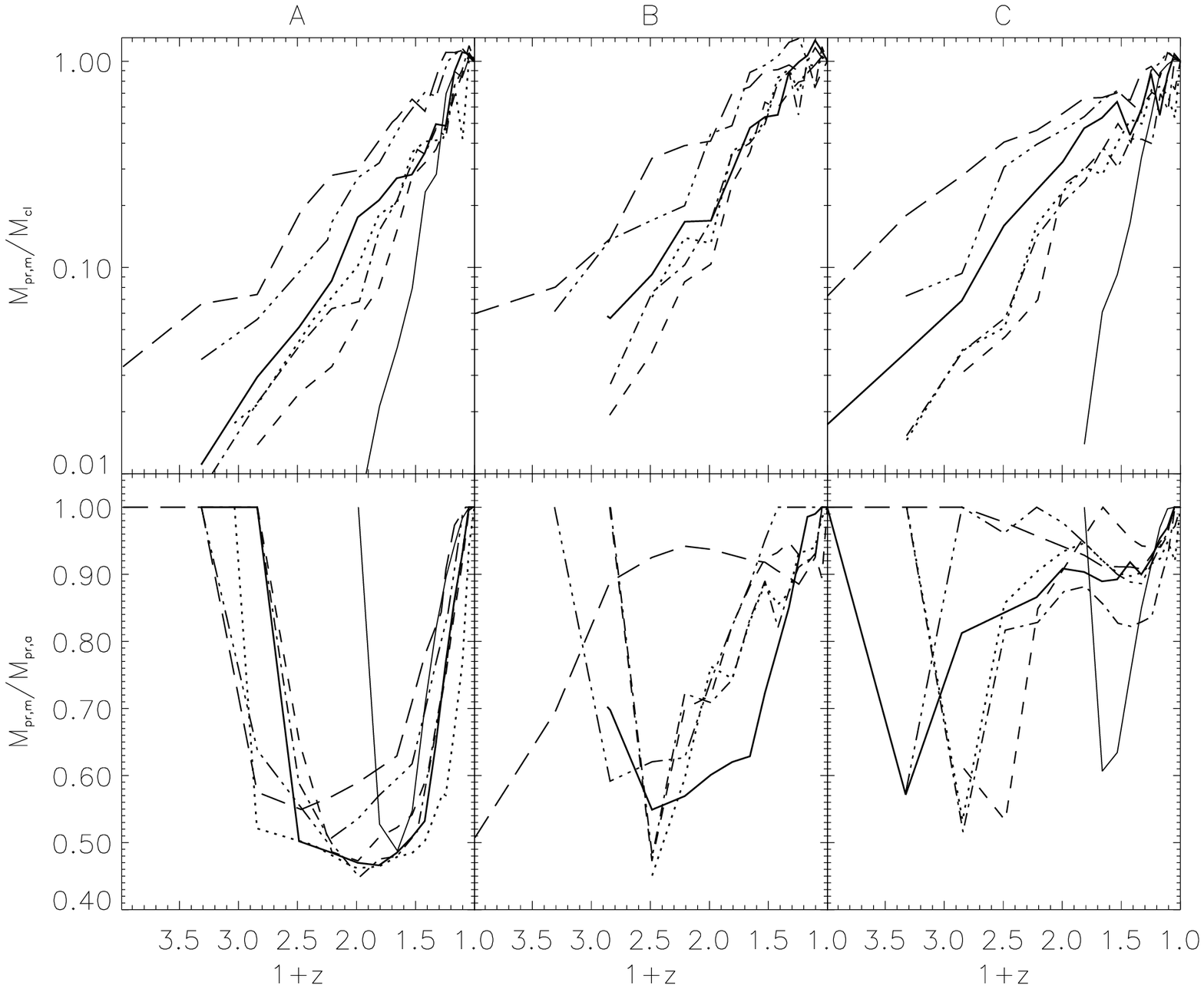}
\caption{a) Mass fraction of the most massive progenitor to the total
cluster mass and b) ratio of the mass of the most massive progenitor to the
mass of all gravitational bound sub-clumps as function of redshift.  As in
figure~\ref{figps} the solid line corresponds to CDM,the dash
triple-dotted to LCDM, the long-dashed to OCDM, the thick dotted, dash dotted and
dashed line to CHDM I, II and III, and the thin dotted line to the HDM model.}
\label{figmbn}
%\end{center}
\end{figure*}

This visual impression can be quantified by looking at the mass ratios of the
progenitors of the $z=0$ cluster.  The progentiors of a halo are found using the
halo identification techniques described in section \ref{IHS} for every
output. Each collapsed object for which more than half of its particles end up
in the cluster at $z = 0$, defines a progenitor.  Figure~\ref{figmbn}a shows the
mass ratio of the most massive progenitor $M_{pr,max}$, to the total cluster
mass $M_{cl}$ at $z = 0$ as a function of redshift.  This ratio represents the
nonlinear growth of the cluster.  Consistent with the expectation of linear
theory, the low $\Omega$ models exhibit a faster growth of structure at high
redshifts as compared to the $\Omega=1$ CDM models. The growth rate of mixed
dark matter models is somewhat suppressed at higher redshifts due to the
influence of the hot component.
%({following is discarded, because there is an $M\propto (Exp(1+z))^{-(1--3)}$
%behavior.})

A good measure for the merging history of a cluster is the ratio of the mass of
the most massive progenitor $M_{pr,max}$, to the mass of all (collapsed)
progenitors $M_{pr,all}$. If a halo grows only by accumulating (not collapsed)
infalling matter this ratio would be one at all times. Departures from one are a
sign of mergers. The degree of departure is governed by the number and mass
ratio of the progenitors and thus to the merger rate. Figure~\ref{figmbn} b)
shows the ratio $M_{pr,max}/M_{pr,all}$ for our clusters.

While the redshift dependence of the cluster mass is very similar for the
different realisations, its merging history is very different
(figure~\ref{figmbn}b).  For realisation A, the ratio $M_{pr,max}/M_{pr,all}$
lies close to its minimum for a long time and the most massive progenitor
comprises only about 50-60\% of the mass in all collapsed objects. This reflects
the visual impression from figure~\ref{figala} that the cluster has several
progenitors of similar mass which merge together at a redshift close to
zero. Realisation B and especially C exhibit much less merging events and the
total amount of mass in collapsed structures is largely dominated by the most
massive progenitor, consistent with figure~\ref{figala} which shows for these
realisations a dominant mass concentration near the center. The differences in
the merging histories between differing cosmological models are much less than
those between different realisations and they are barely significant. Such a
similarity in the evolution pattern is also predicted by the modified
Press-Schechter model (Bond \etal 1991, Lacey \& Cole 1993), which predict
identical merging trees (number of mergers) for Gaussian fields with identical
phases. Different cosmological models differ primarily in the different
assignment of an individual merging event with a physical time coordinate, i.e.
the differing redshift dependence of $M_*(z)$.

\subsection{Shape and angular momentum of the haloes}

\begin{figure*}
\centering
\epsfxsize= \hsize\epsffile{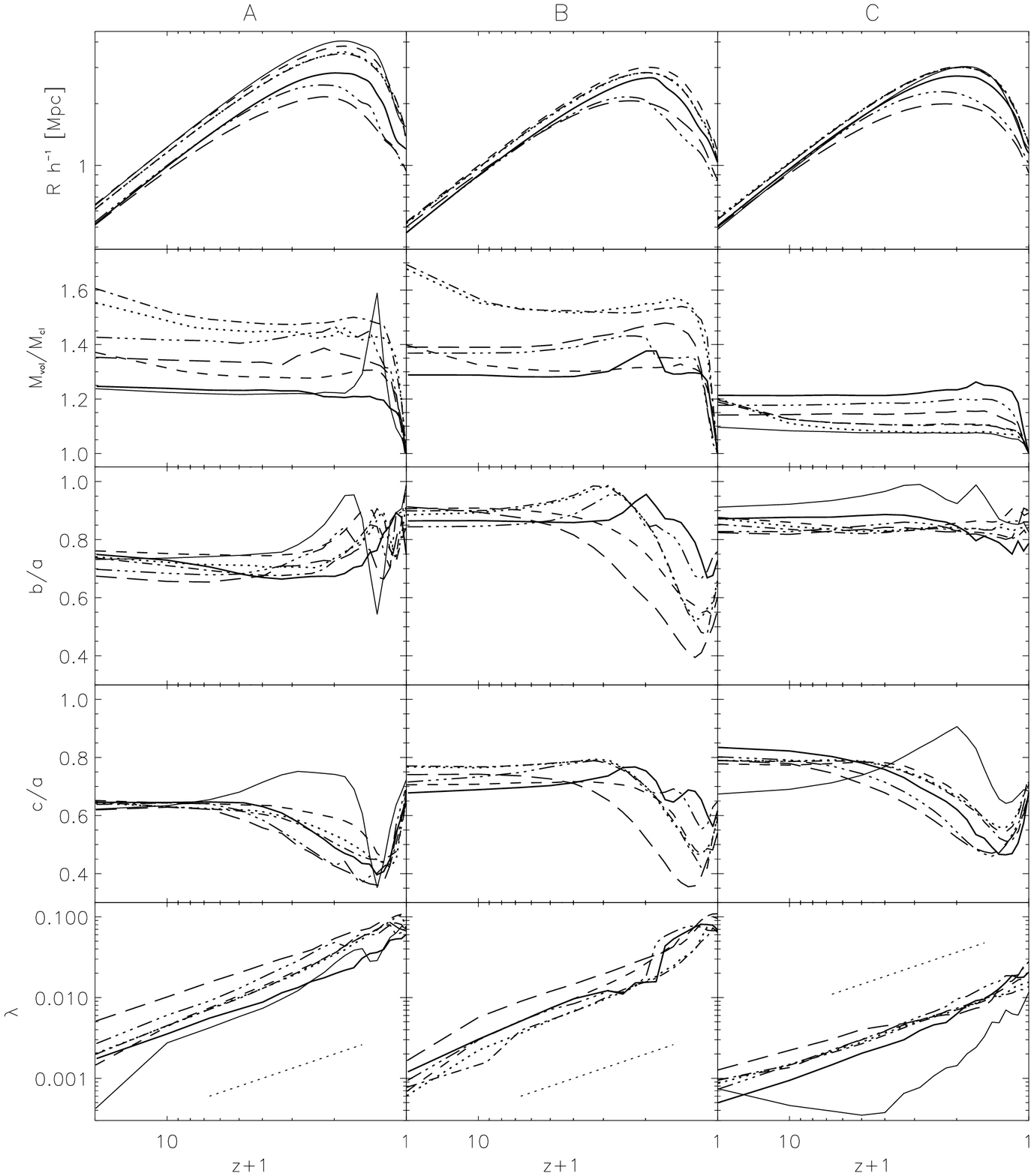}
\caption{Time evolution of the size $R$, the mass ratio
$M_{vol}/M_{cl}$, the axes ratio $b/a$ and $c/a$ and the spin parameter
$\lambda$ of the region occupied by the particles which end up in a halo at $z =
0$. The CDM clusters correspond to the solid lines, LCDM clusters to the
dash triple-dotted lines, OCDM clusters to long-dashed lines, the clusters of
the three CHDM models to the thick dotted (I), dash dotted (II) and dashed (III) and
the HDM clusters to the thin dotted lines. The straight dotted line indicates the growth
of $\lambda$ proportional to $(1+z)^{-1}$. } 
\label{figtpt}
%\end{center}
\end{figure*}

Further aspects of cluster collapse can be studied by investigating the mass
distribution within the volume ${\mathcal V}(z)$, defined as the volume covered
by those particles which at $z=0$ lie within the virial radius $r_{\rm vir}$ of
the cluster (hereafter the cluster particles).  The shape of ${\mathcal V}(z)$
deviates from a sphere at higher redshifts. We define the boundaries of
${\mathcal V}(z)$ by those cluster particles which have the  largest
distance to the center of mass of all cluster particles. To identify
the boundary particles the volume around this center of mass is
devided in its eight octants. Inside each octant the absolute values
of the three space coordinates of each cluster particle are compared
to the absolute coordinates of all other cluster particles. A particle
lies inside ${\mathcal V}(z)$, if one of the remaining particles has a
larger absolute coordinate in one direction than the reference
particle.  With the same method one
can also select all particles inside ${\mathcal V}(z)$ by comparing 
the coordinates of all particles with those of the boundary particles.
We then approximate ${\mathcal V}(z)$ by its ellipsoid of inertia (principal axes
$a$, $b$, $c$) and define the size $R$ of the volume ${\mathcal V}(z)$ as the
radius of a sphere with the same volume as the ellipsoid of inertia of
${\mathcal V}(z)$. $M_{\rm vol}$ is the mass enclosed by ${\mathcal V}(z)$. The
angular momentum $L_{\rm vol}$ of the cluster region can be parametrised using
the spin parameter $\lambda$. It can be generally defined as
\begin{equation}
\lambda = \frac{L_{\rm vol}}{G^{\frac{1}{2}} M_{\rm vol}^{\frac{3}{2}}
R^{\frac{1}{2}}}.
\label{lam}
\end{equation}
This definition of $\lambda$ also works if the cluster region as a whole is not
gravitionally bound.  For virialized objects $G\,M_{\rm vol}/R$ gives roughly
the total energy $E$ of the object thus leading to the usual definition of the
spin parameter, $\lambda = L_{\rm vol}\sqrt{|E|}/(G\,M^{2.5})$.

In figure~\ref{figtpt} the time evolution of the size $R$, of the axis ratios
$b/a$ and $c/a$, of $M_{\rm vol}/M_{\rm cl}$ and of $\lambda$ is shown. These
figures demonstrate again, that for hierarchical models the general picture of
collapse is very similar within the same realisation, nearly independent of the
cosmogony. Only the HDM clusters show departures from this picture.

\begin{figure}
\centering
\epsfxsize= \hsize\epsffile{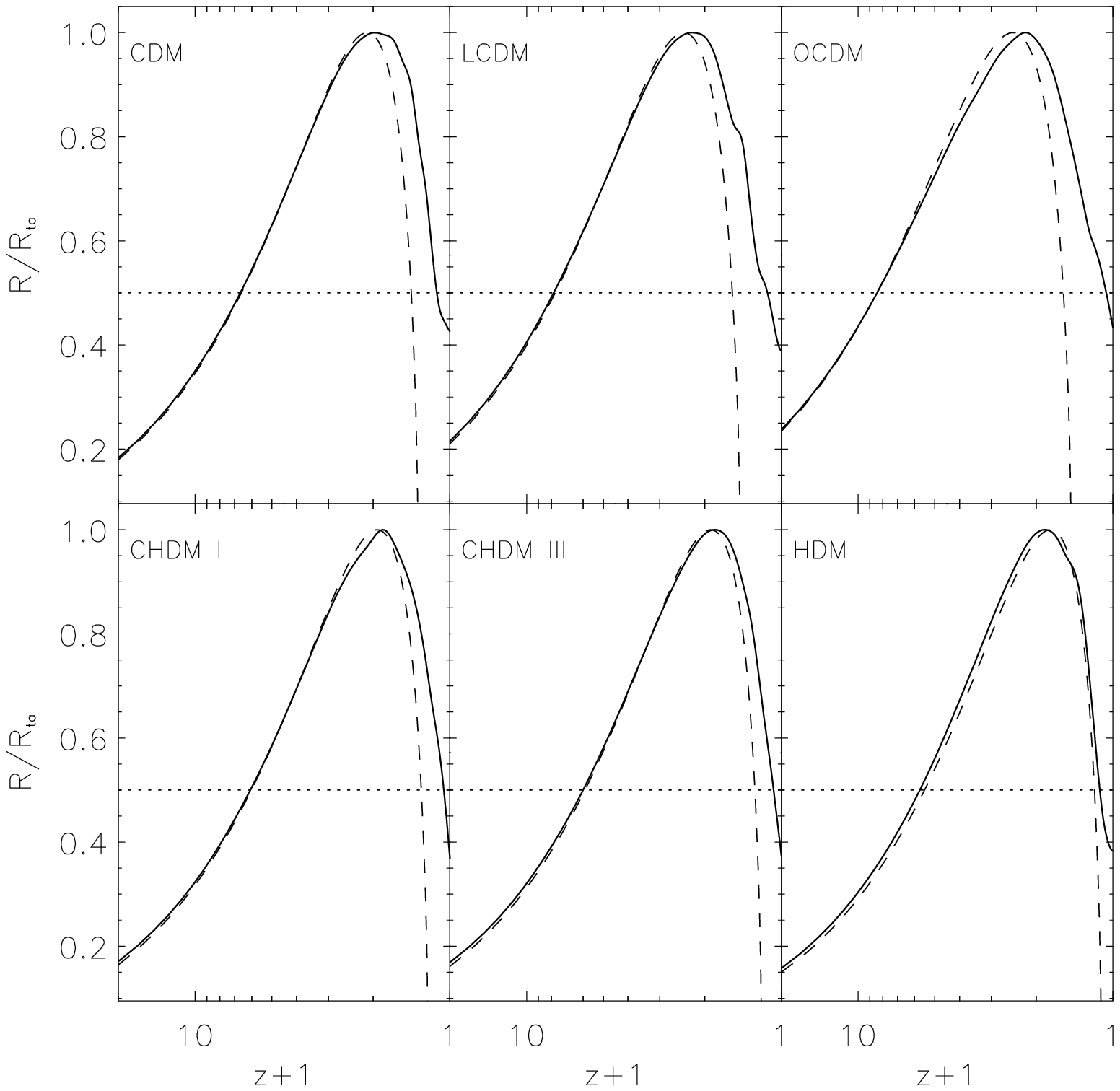}
\caption{Time evolution of the size $R$ of all haloes of realisation A
(apart from the CHDM II model) compared to the time evolution of the spherical
top-hat model of a sphere with the same initial mass and turnaround radius.}
\label{figtp3}
%\end{center}
\end{figure}

In figure~\ref{figtp3} $R(z)$ scaled with the turnaround radius $R_{ta}$ is
explicitly compared to the solution of the top-hat model for the clusters of
realisation A. For each halo the Tolman-Bondi equation~(\ref{TBdi}) is solved
using the mass inside ${\mathcal V}(z)$ at $z = 20$ as the collapsing mass
$M$. The energy $E$ is derived at turnaround, $E = G M R_{ta}^{-1} + \Lambda/6
R_{ta}^2$.  Before turnaround all models show a very good agreement between the
numerical simulation and the spherical top hat model .  After turnaround the
analytical solution gets singular whereas simulated clusters virialise. In all
cases the virial radius is approximately 40 \% of the turnaround radius. But due
to the definition of $R$ as the effective radius of the ellipsoid of inertia it
also reflects the mass distribution inside the halo region. The strongly
clustered matter distribution of the haloes leads to a smaller size than a
homogeneous one of the same mass as assumed in the top-hat model. It is
remarkable that the deviations from the top-hat solutions is smallest for HDM
model. As discussed later the evolution of the HDM clusters is less violent than
for the other models enabling an almost undisturbed infall of matter.

The collapse is initially driven by a mass which is $20-50\%$ larger than the
halo mass at $z=0$. The deviation is least pronounced for realisation C which
was shown to have a merging history most similar to a spherical accretion model.
The surplus of mass is reduced at low redshift (between $z \approx 0.4$ and 0.1)
when the collapse along the smallest axis of the ellipsoid of inertia has been
completed and the axis ratio $c/a$ reaches its minimum.  The lost mass is partly
linked with particles which gain enough energy in the violent phase after
collapse of the smallest axis to leave the halo.  The residual mass forms small
clumps which lie initially within ${\cal V}(z)$ (preferably at the end of the
major axis of the ellipsoid of inertia) but do not participate in the
collapse. In the final state of the halo collapse they are not enclosed by
${\cal V}(z)$ anymore.

The evolution of the axis ratio of the ellipsoid of inertia 
shows the anisotropy
of the collapse. It depends strongly on the local realisation of the initial
conditions. In realisations A and C the collapse is faster only along 
the smallest
axis, while the two longer axes remain nearly equal. In realization B
on the other hand both small axes collapse faster than the major axis. 
As discussed in section~\ref{mma} the clusters of
realisation A are formed out of a very filamentary structures where all
filaments lie in one sheet. Here the collapse perpendicular to the sheet
(corresponding to the axis c) is faster than the merging of the clumps forming
in the filaments. The same happens for the clusters of realisation C. The
initial structure is less filamentary but forms a sheet. A different behaviour
is observed for the realisation B, since here the clusters evolve out of two
major objects forming one filament. The attraction of these 
two components needs
more time than the collapse of each component alone.

The evolution of the spin parameter $\lambda$ is almost unaffected by the mass
loss and the anisotropy of the collapse. It grows approximately proportional to
the expansion factor $a = (1+z)^{-1}$, the same rate expected in linear
theory. According to linear theory, the total angular momentum should grow like
$a^2\dot D$ (White 1984, Steinmetz \& Bartelmann 1995). Before turn around, $R
\propto 1+z$, and therefore from the definition of $\lambda$ in equation
(\ref{lam}) it grows to first order as $\lambda\propto a^{3/2}\dot D$.  At high
redshifts the Einstein-de Sitter relations $D \propto a$ and $a\propto t^{2/3}$
are approximately valid in all cosmological models lead to the
$\lambda\propto(1+z)^{-1}$ dependence. Interestingly, as shown in
figure~\ref{figtpt} nearly all clusters follow this growth even at low redshift.
The final values of $\lambda$ lie in the range $0.06 - 0.1$ for realisations A
and B, and $0.01 - 0.02$ for realisation C. Several massive clumps are located
in the neighbourhood of the clusters which form in realisation A and B causing
relatively strong tidal fields. The shape of ${\mathcal V}(z)$ is quite elongated
and thus ${\mathcal V}(z)$ possesses a high quadrupole moment which can be
efficiently spun up by the tidal field.  Realisation C on the other hand has no
nearby massive object and the shape of ${\mathcal V}(z)$ is nearly
spherical. The weak tidal effects can work only inefficiently and the resulting
spin up of the halo is therefore small (White 1984; Steinmetz and Bartelmann
1995). Thus we find that the amplitude of $\lambda$ can vary significantly
between different clusters in a given cosmogony, while the difference between
different cosmogonies are again small.

The formation of the two HDM clusters is qualitatively different from the
hierarchically clustering models. No bound objects are formed until $z \approx
1$, but the matter accumulates in a sheet, which lies roughly perpendicular to
the major axis of ${\mathcal V}(z_{\rm init})$.  The different orientation of
the sheet and of ${\mathcal V(z_{\rm init})}$ is compensated during
the collapse. A more spherical ellipsoid of inertia develops before the collapse
along the smallest axis has finished. Therefore, the axis ratio of the ellipsoid
of inertia increase at high redshifts (figure
\ref{figtpt}). The evolution of the spin parameter is much slower for HDM C.
Due to the absence of nearby overdensities and the related tidal
fields, angular momentum does not grow at high redshifts.

\section{The structure of clusters}

\subsection{Cluster profiles at $z = 0$} \label{CP0}

As shown in figures ~\ref{figala} and \ref{figalb}, galaxy clusters build up
quite irregularly. The formation can be characterized by mergers of large clumps
as in the case of realization A or by infall of less massive objects as in
realization C.  A smooth spheroidal matter distribution can only be observed
near the center of the cluster, where the particle distribution has
virialised. But even at low redshift matter infall and tidal fields are still
present, especially in the case of the high $\Omega$ models.

We fit an NFW profile to the binned radial matter distribution. Fit parameters
are determined using the the circular velocity $v_{\rm cir}(r)$ since it is less
noisy. The circular velocity $v_{\rm cir,n}(r)$ of an NFW profile is given by
\begin{equation}
v_{\rm cir,n}^2(x) = 4 \pi G \varrho_b \delta_n a_n^2 x^{-1} \left[ \ln(1 + x) -
\frac{x}{1 + x}\right]
\label{nvcn}
\end{equation}
with $x=r/a_n$. The circular velocity at the virial radius is $v_{\rm vir} =
v_{\rm cir,n}(c) =
\sqrt{G M_{\rm vir}/r_{\rm vir}}$ with $c = r_{\rm vir}/a_n$. Therefore
$\delta_n$ is also a function of $a_n$ and the density profile depends only on
one free parameter (Navarro
\etal 1996).

The scale length $a_n$ is determined by means of a nonlinear least square fit
weighted by the errors in the circular velocity for each bin. Errors are
determined by means of a bootstrap method (Efron 1979) using a resampling rate
of $100$.  Each sample is binned using the same bin boundaries as the original
particle distribution. Therefore the particle number and the mass within one bin
varies and allows one to estimate the standard deviation in $v_{\rm
cir}$. Calculating the bin error using the bootstrap method takes into account
not only the discretisation error ($\propto 1/\sqrt{N_{bin}}$) but also the
deviation of the halo region from a sphere.  In table~\ref{tb3} all profile
parameters of a halo are listed. We also give an error estimate for the
characteristic radius $a_n$ derived form the bootstrap analysis. This error
typically is of the order of 10 per cent.  For CHDM clusters it is little bit
smaller since the bins contain more particles.

\begin{table*}
\caption{\label{tb3}Profile parameters for the 20 considered clusters. Listed
are the
viral mass $M_{\rm vir}$, the virial radius $r_{\rm vir}$, the scale length
$a_n$ with
its 3$\sigma$ bootstrap error, the characteristic overdensity $\delta_n/4$ and
the concentration $c = r_{\rm vir}/a_n$. (The number in the brackets give the
error
in $a_n$ as a percentage of $a_n$.)}
\centering
\begin{tabular}{lccccc}
Model & $M_{\rm vir}\,[10^{15}\;h^{-1}\;\Omega_0\;M_{\odot}]$ & $r_{\rm vir}\,
[h^{-1}\;{\rm Mpc}]$  & $a_n\, [h^{-1}\;{\rm Mpc}]$ & $\delta_n$ &  $c$
\\ \hline
OCDM A &  1.01 &  1.30 & $0.23 \pm 0.03$ (12\%) &   5950  &  5.76 \\
OCDM B &  0.91 &  1.25 & $0.18 \pm 0.02$ (10\%) &  10150  &  7.28  \\
OCDM C &  1.12 &  1.34 & $0.18 \pm 0.01$ (7\%) &  11550  &  7.63  \\
LCDM A & 0.95 &  1.34 & $0.21 \pm 0.03$ (12\%) &  7010  &  6.65  \\
LCDM B & 0.69 &  1.21 & $0.20 \pm 0.03$ (14\%) &   5820  &  6.12 \\
LCDM C &  1.11 &  1.42 & $0.25 \pm 0.03$ (14\%) &  4840  &  5.71 \\
SCDM A &  1.01 &  1.69 & $0.42 \pm 0.12$ (29\%) &  1260  &  4.11 \\
SCDM B & 0.71 &  1.50 & $0.29 \pm 0.05$ (16\%) &  2250  &  5.33 \\
SCDM C & 0.96 &  1.66 & $0.28 \pm 0.07$ (26\%) &  3180  &  6.19 \\
CHDM I A &  1.31 &  1.84 & $0.41 \pm 0.05$ (12\%) &  1530  &  4.52 \\
CHDM I B & 0.76 &  1.54 & $0.30 \pm 0.04$ (15\%) &  2150  &  5.20 \\
CHDM I C &  1.32 &  1.85 & $0.35 \pm 0.03$ (5\%) &  2360  &  5.48 \\
CHDM II A &  1.24 &  1.81 & $0.49 \pm 0.04$ (8\%) &  1030  &  3.76 \\
CHDM II B & 0.79 &  1.55 & $0.30 \pm 0.03$ (11\%) &  2250  &  5.31 \\
CHDM II C &  1.27 &  1.82 & $0.30 \pm 0.02$ (8\%) &  3280  &  6.32 \\
CHDM III A &  1.70 &  2.01 & $0.63 \pm 0.10$ (15\%) &  750  &  3.23 \\
CHDM III B & 0.99 &  1.66 & $0.46 \pm 0.05$ (10\%) &   1010  &  3.73 \\
CHDM III C &  1.26 &  1.82 & $0.40 \pm 0.02$ (6\%) &  1650  &  4.66 \\
HDM A &  1.78 &  2.04 & $0.85 \pm 0.12$ (14\%) &  420 &  2.46 \\
HDM C & 0.96 &  1.66 & $0.45 \pm 0.19$ (43\%) &  1070 &  3.83 \\
\end{tabular}
%\end{center}
\vspace{0.1 cm}
\end{table*}

\begin{figure*}
\hskip0.05\hsize\epsfxsize=0.9\hsize\epsffile{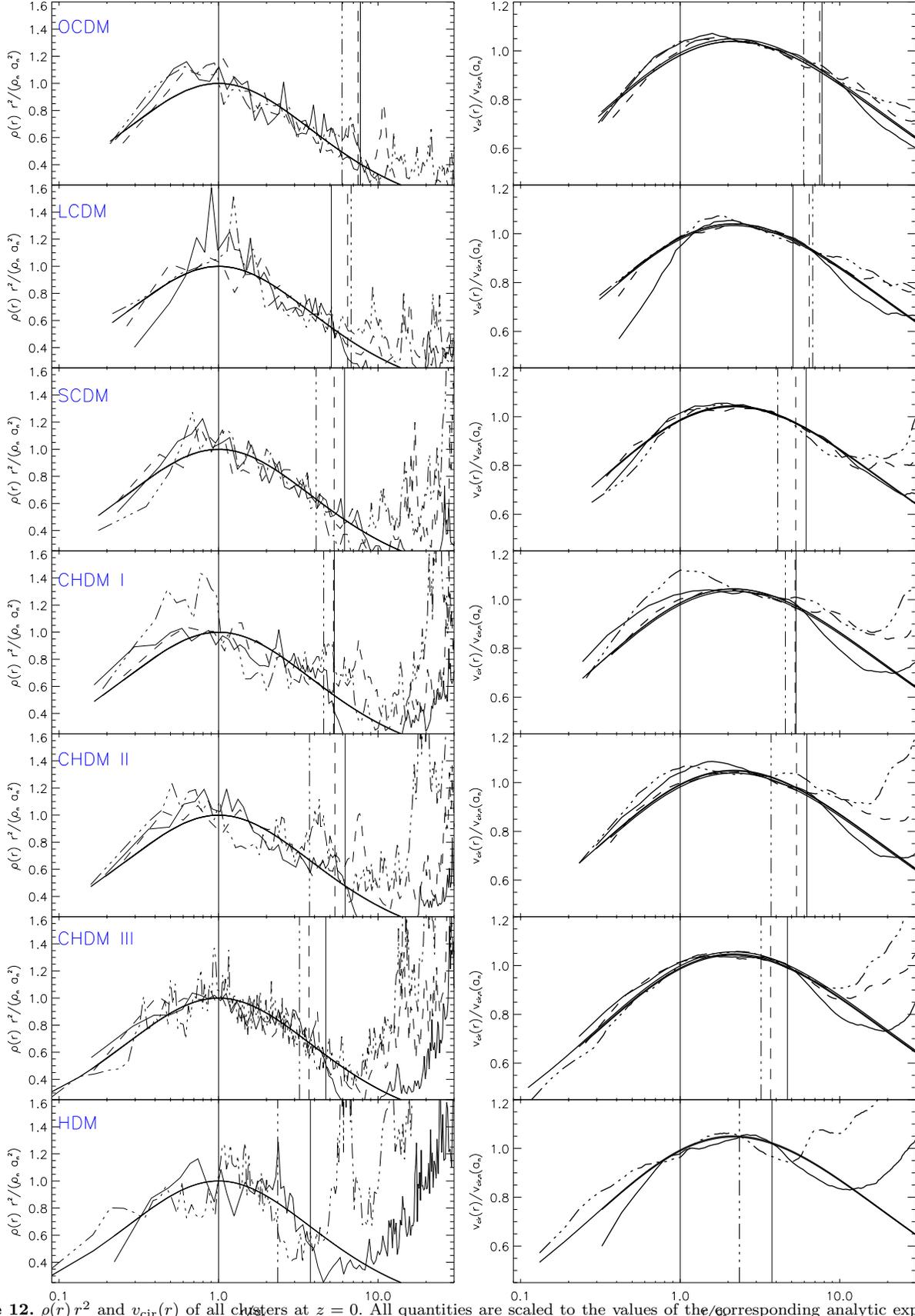}
\caption{$\varrho(r)\,r^2$ and $v_{\rm cir}(r)$ of all clusters at
$z = 0$. All quantities are scaled to the values of the corresponding analytic
expression at the scale radius $a_n$. The dash dotted, dashed and the
long-dashed lines represent realizations A, B and C, respectively. The solid
line gives the analytic expression using the parameters of the best fit to the
NFW profile.}
%The vertical bars on the right mark the virial radii of the
%clusters, the arrows indicate the softening length.
\label{figprfa}
%\end{center}
\end{figure*}

In figure~\ref{figprfa}, $r^2\varrho(r)$ and the circular velocity $v_{\rm
cir}(r)$ are shown as function of $x=r/a_n$ for all clusters at $z =
0$. $r^2\varrho$ and $v_{\rm cir}$ are scaled to the value of the analytical fit
at the scale radius. This enables a fair comparison of the haloes since under
such a scaling all profiles which are fitted well by equation $\ref{nfw}$ should
have identical shapes. The solid lines in figure~\ref{figprfa} represent these
reference profiles based on the NFW profile. All profiles are very well fitted
by the NFW profile up to the virial radius. Larger deviations especially in form
of spikes are produced by infalling subclumps (Tormen 1996), or due to
incomplete relaxation as in the case of the HDM A. Both effects lead also to a
larger error in $a_n$.

\begin{figure*}
\centering
\hskip0.05\hsize\epsfxsize=0.9\hsize\epsffile{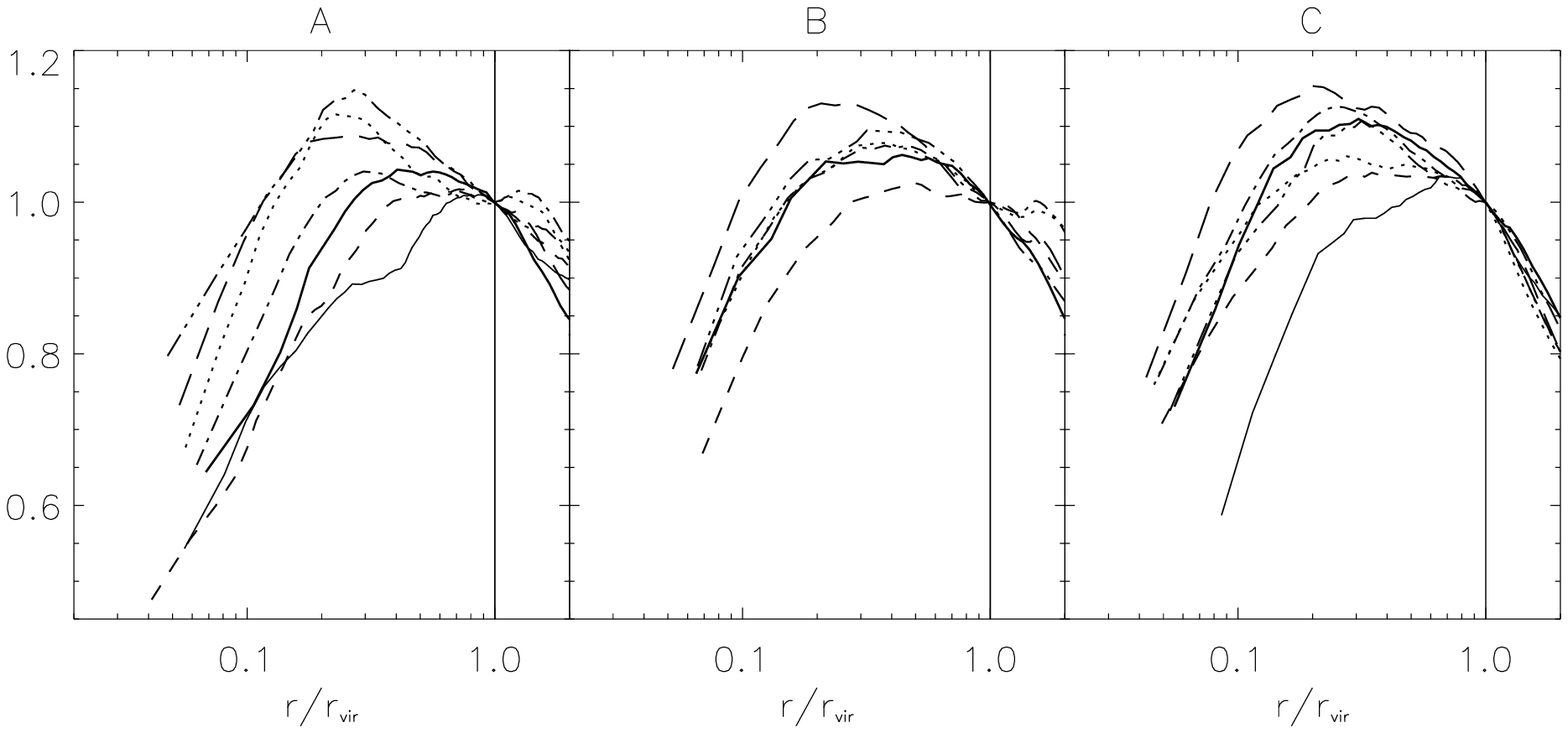}
\caption{The profile of $v_{\rm cir}(r)$ is compared for all clusters of one
realization at $z = 0$. Now the distance is scaled with the virial radius
$r_{\rm vir}$ and $v_{\rm cir}(r)$ with $v_{\rm vir}$. The lines are the same as
in figure~\ref{figps}.}
\label{figprfb}
%\end{center}
\end{figure*}

The general shape of the haloes follows the NFW profile is independent of the
cosmogony. The only difference between the models is the concentration $c =
r_{\rm vir}/a_n$ of haloes, characterized by the ratio of the viral radius to
the scale length.  $c$ is given in table~\ref{tb3} for all haloes. It can also
be directly read off figure~\ref{figprfa} looking at the x-values of the virial
radii. A small value of $c$ means that the density in the center is lower
compared to haloes with a high $c$. This is shown in figure~\ref{figprfb}, where
the profiles of the circular velocity of the clusters of one realization are
compared.  For pure cold dark matter, there is a trend towards higher
concentration from the CDM model to the LCDM and OCDM model.  This trend can be
well understood since the characteristic overdensity $\delta_n
\propto (1+z_f)^3$ (Navarro \etal 1996b) where $z_f$ is the formation time of
the halo.  
In $\Omega<1$ universes structures form at an earlier epoch where the density of
the universe is higher. The trend is most pronounced for the OCDM model, since
in this scenario, structures form at the highest redshifts. There is also a
strong correlation of $c$ with increasing influence of the hot component. The
concentration becomes smaller as the hot dark matter becomes more dominant.
Structure formation is retarded to very low redshifts.

\subsection{The asymptotic behaviour of the density profiles at large radii}

Both the NFW and the Hernquist profile have been shown to provide a fairly
accurate fit to the radial matter distribution of haloes formed in
hierarchically clustering scenarios, though the NFW profile seem to provide a
slightly better fit, especially at large radii. We investigate more
quantitatively the asymptotic behavior at large radii using a generalized form
of the NFW profile:
\begin{eqnarray}
\frac{\varrho(r)}{\varrho_b} & = & \delta_g \frac{a_g}{r (1 +
\frac{r}{a_g})^m}\, ,
\label{gen} \\
v_{\rm cir,g}^2(x) & = & 4 \pi G \varrho_b \delta_g a_g^2 x^{-1} \nonumber
\times \\ & &
\left[\frac{1 + m x - x^2 + m x^2 - (1+x)^m}{(2 - 3 m + m^2)(1+x)^m} \right]
\label{nvcf}.
\end{eqnarray}
For $m = 2$ the expression~(\ref{nvcn}) has to be used for $ v_{\rm
cir,g}^2(x)$.  We have thus introduced an additional free parameter $m$ which
sets the asymptotic logarithmic slope $\alpha=m+1$ of the profile at large
radii.  $m = 2$ and $m = 3$ correspond to the NFW and the Hernquist profile,
respectively.  The physical meaning of the other two parameters, $a_g$ and
$\delta_g$ is the same as that of $a_n$ and $\delta_n$ in the NFW profile,
namely characteristic scale radius and central density.

By fitting equation (\ref{gen}) insidue $r_{vir}$ to the measured profiles for
all available haloes we get the distribution of $m$. To enlarge the halo sample
we include the most massive progenitor of each cluster identified in the outputs
after $z = 0.5$. With our sampling rate of one output per $\Delta a= 0.05$ we
get a total of $193$ haloes covering a mass range of $1.2
\times 10^{14} h^{-1} \Omega M_{\odot}$ to $1.9 \times 10^{15} h^{-1} \Omega
M_{\odot}$. The halo sample of each cluster realization is not strictly
independent as these haloes belong to the same evolution path. Due to the chosen
sampling rate, however, the haloes are identified at different states of the
evolution of a cluster.  Each halo has included different amount of
nonlinearities and represents a different state of relaxation.  Furthermore, the
asymptotic slope probes the outer part of a halo, which is mainly built up from
the newly infalling matter. Though not strictly independent, our enlarged sample
can be considered to represent the typical amount of scatter in the asymptotic
behaviour at large radii.

\begin{figure}
\centering
\epsfxsize= \hsize\epsffile{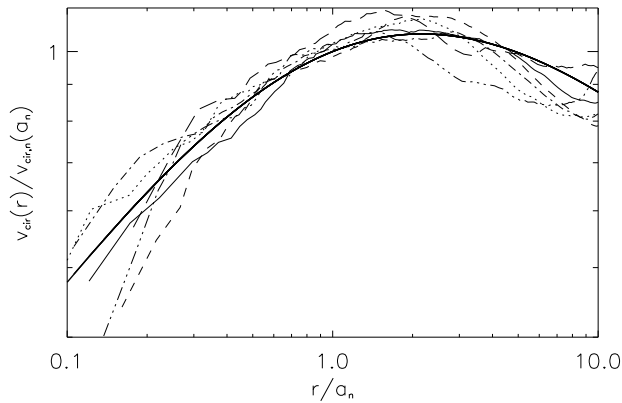}
\caption{$v_{\rm cir}$ of the CDM A cluster at $z = 0$
(solid line) and of its most massive progenitors at $z = 0.05$ (dotted line),
$0.10$ (dashed line), $0.17$ (dash-dotted line), $0.24$ (dash triple-dotted
line) and $0.33$ long-dashed line). The smooth thick solid line correspond to
the NFW profile.}
\label{figtprof}
\end{figure}

The radial density distribution is similar in all of these objects as can be
seen in figure~\ref{figtprof}. Here we plot the profiles of $v_{\rm cir}$ of the
CDM A cluster at $z =0$ and of its most massive progenitors at seven earlier
redshifts.  The general shape of the profiles is comparable to the NFW profile,
independent of redshift.

\begin{figure}
\centering
\epsfxsize= \hsize\epsffile{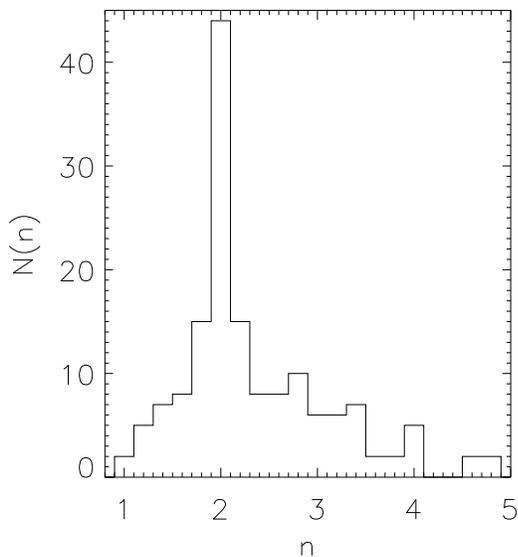}
\caption{Number distribution $N(m)$ for the asymptotic slope $m$ of the
density profiles of all clusters and their most massive progenitors.}
\label{fignv}
%\end{center}
\end{figure}

In figure~\ref{fignv} the number distribution $N(m)$ of the asymptotic slope of
the profiles is plotted. The distribution is strongly peaked at $m = 2$, the
value of the NFW profile.  This peak is a common feature for all cosmologies.
There is no relation between $N(m)$ and redshift.  We are thus led to conclude
that the NFW profile has the best fit asymptotic slope at large $r$, for a
generalized profile of the form of equation (\ref{gen}).

A different asymptotic behavior ($m\ne 2$) can be observed for haloes which
participate in a merging event. Merging subclumps change the matter distribution
along the infall direction, which is visible as spikes in the radial profile as
mention earlier.  Depending on the distance of the subclump relative to the halo
center these spikes affect the asymptotic slope at large radii. A flattening of
the profile results if a subclump is at radii slightly larger than $r_{\rm
vir}$; on the other hand shortly after a subclump crosses $r_{\rm vir}$, the
resultant spike leads to a steeper outer profile.

\subsection{Non-radial motions and the evolution of the density profile}
\label{AAE}

In the previous subsections we have demonstrated the universality of the
measured density profiles of clusters.  For $r<r_{\rm vir}$, the flattening
below a certain scale length and the asymptotic decay of the profiles are well
described for more or less relaxed haloes by the NFW profile. This general form
of the radial density distribution in a halo is independent of redshift and the
cosmological model. The evolution path of a single halo also does not appear to
affect the profile.  Recently Syer \& White (1996) have argued that the inner
slope of the universal density profile arises from repeated mergers. The
clusters we have simulated however have very different merging
histories; indeed
the first collapsed objects and the entire haloes of our HDM simulations are
formed without any significant mergers events. In this section we consider
the role of non-radial motions in producing an NFW-like profile.

\begin{figure*}
\centering
\hskip0.05\hsize\epsfxsize=0.9\hsize\epsffile{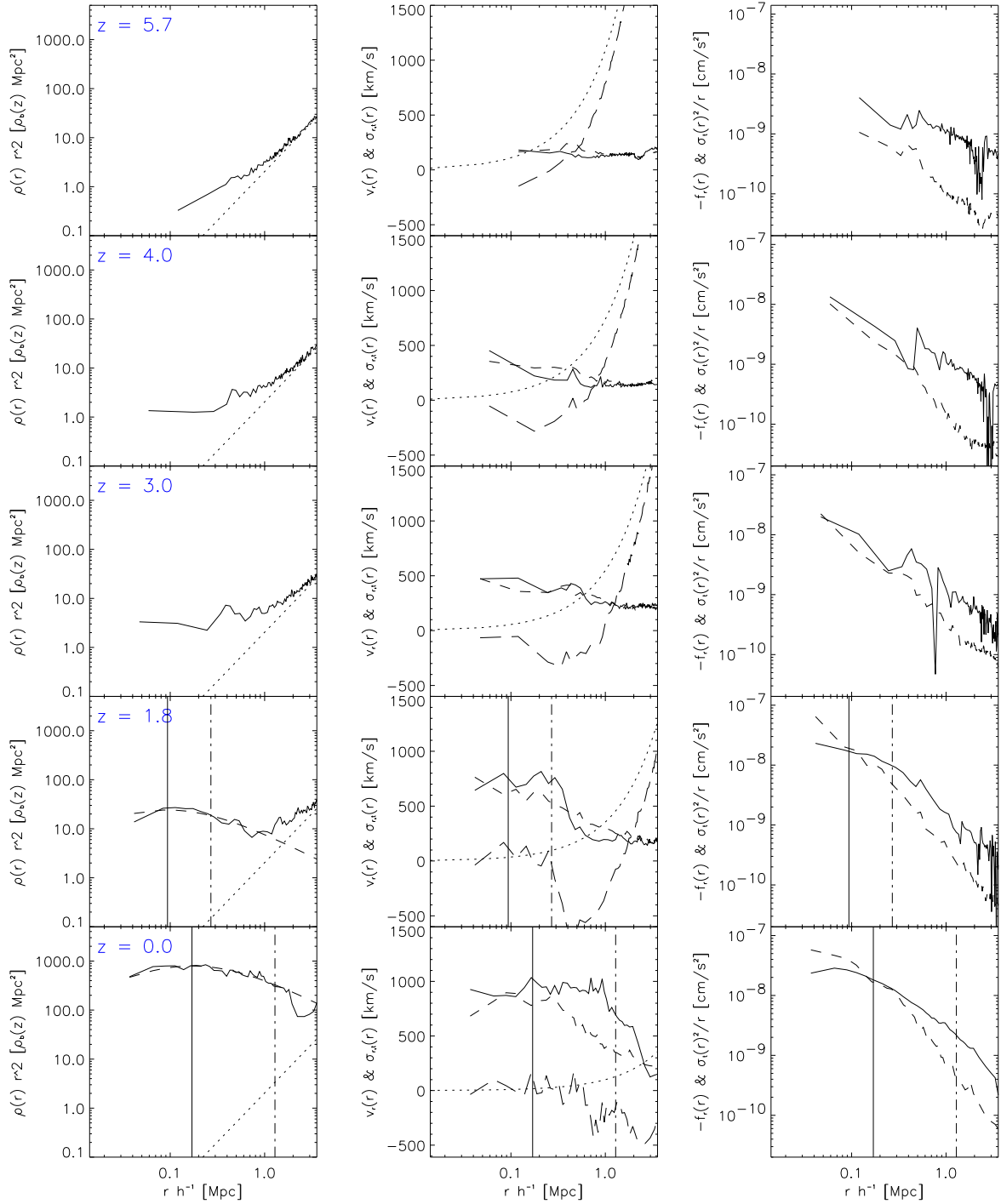}
\caption{\label{figevol}
The formation of an NFW-like density profile is demonstrated with the OCDM C
cluster. In the left panel the profile of $\varrho(r,z) r^2$ is plotted for the
most massive progenitor. The NFW fit is the dashed curve, while the dotted line
is for $\varrho(r)={\rm constant}$.  The middle panel shows the profiles for the
mean radial velocity $v_{\rm r}(r,z)$ (long dashed) and the radial
$\sigma_r(r,z)$ (solid) as well as the transverse $\sigma_t(r,z)$ (dashed)
velocity dispersion.  The Hubble flow is indicated by the dotted line. In the
right panel the negative radial gravitational force $f_r(r,a)$ (solid curve) is
plotted together with the ``centrifugal'' 
force $\sigma_t(r,z)^2/r$ (long dashed
curve). In the lower two rows the upright solid line shows $a_n$ and 
the dash dotted line $r_{\rm vir}$ at the given redshift.}
\end{figure*}

It is illustrative to examine the time evolution of the density
profile in parallel with the velocity structure of the halo as 
measured from the mean radial velocity $v_r(r)$, and the radial 
[$\sigma_r(r)$] and the transverse
[$\sigma_t(r)$] velocity dispersion of particles. 
In figure~\ref{figevol} this evolution is shown for the OCDM C cluster. For
the region where the most massive progenitor is formed, the profile of the
density $\varrho(r) r^2$ is plotted as a function of distance $r$ at different
redshifts (left panels).  The middle panels shows the profiles of 
$v_r(r), \sigma_r(r)$ and $\sigma_t(r,z)$.  The right panel shows the 
profile of $\sigma_t(r,z)^2/r$ in
comparison with the radial component of the gravitational force. 

Figure~\ref{figevol} shows how the development of a density profile 
like the NFW profile of equation \ref{nfw} 
is accompanied by a transition from mainly radial infall of
the halo particles to virialised motions with vanishing mean radial velocity.
At $z = 5.7$ all particles at radii below $0.3 h^{-1}\,$Mpc have turned around
and are falling towards the density maximum. The density profile is
not much steeper
than a constant density profile (the logarithmic slope $\alpha \simeq -0.8$). 
The first particles cross the central region
before $z = 4$ and move outwards again. Due to the overlap of 
the particles crossing the center for the first time and those
which have fallen in and out several times, the density increases in the center
as described by the secondary infall model (e.g. Bertschinger 1985).  This
results in the  slope $\alpha$ becoming lower than $-2$
at radii below $0.2 h^{-1} {\mathrm Mpc}$. 
At $z\approx 3$ a large clump starts to merge with the progenitor, as
evident by the spike in the density profile at $0.4 h^{-1}$Mpc. 
By $z\simeq 2$ the slope has again become shallower than $-2$ in the center
while it remains steeper than $-2$ for $a_n<r<r_{\rm vir}$. 
The shape of the density profile below $r_{\rm vir}$ remains almost 
stable up to $z = 0$. Merging events temporarily disturb
this matter arrangement, but the density distribution inside always
returns to the NFW form. 

The velocity structure near the center changes during collapse as
shown in the middle panels of figure~\ref{figevol}.  Due to the
mixing of inward and outward moving particles the mean radial velocity
$v_r(r)$ in the
central region approaches zero at $z = 4$. After the first relaxed
object has been formed (at $z \simeq 3$),
$v_r(r)$ fluctuates around zero for $r < r_{\rm vir}$.
The radial velocity dispersion
$\sigma_r(r)$ increases with time at early times, but at late times
($z< 1$) remains constant at nearly
$1000$ km/s for all $r<r_{\rm vir}$.  The transverse velocity
dispersion $\sigma_t(r)$ remains approximately equal to
$\sigma_r(r)$, though at late times it remains below $\sigma_r(r)$
in the outer parts of the halo. 
A ``centrifugal'' force, $f_c(r) = \sigma_t^2(r)/r$ can be associated
with the transverse velocities. 
At $z = 5.7$, the radial component of the gravitational force, 
$f_{r}(r)$ is approximately ten times larger in the center than
$f_c(r)$,  but it does not grow significantly (see the right panels
of figure~\ref{figevol}). 
It can be approximated by $- G \bar
\varrho(r,z)\, r$, with the central mean density $\bar \varrho(r,z)$ remaining
almost constant. By $z \simeq 2$, $f_c(r)$ becomes of the same order as
$-f_{r}(r)$ for $r < a_n$, and it becomes larger 
at later times. A close examination of the redshift interval 
between $z=2$ and $z=0$ (not shown in the figure) reveals that there
is very little variation in time in the profiles of $\sigma_t^2/r$ and
$f_r(r)$: once the smooth profiles with a turnover at $r\simeq a_n$
are established they remain the same up to $z=0$. 

The tangential velocities cause particles to move on orbits which can be
locally described as Keplerian ellipses or hyperbolas. 
Typical particle trajectories do not pass
close to the center if $f_{c}(r)$ is of the same order as $f_{r}(r)$ or
larger, and if the transverse velocity of the particles is of 
order the radial velocity. 
Since particles at a given radius do not subsequently 
penetrate into the
inner region of the halo they do not cause the density profile to
steepen as much as it would if they were on radial orbits.  

One can relate the orbit structure of halo particles to the 
density profile that arises in idealised models of secondary infall.
The profile predicted by the purely radial infall model 
(Fillmore \& Goldreich 1984;  Bertschinger 1985) is 
close to what is observed in the outer parts of our haloes. 
In the inner parts however the profile is flattened owing to the 
significant tangential velocities, as described above. This
flattening is consistent with the results of White \& Zaritsky (1992) who
included angular momentum into an infall model. 
They showed that whereas for purely
radial infall, the profile does not become shallower than $\alpha=-2$, the
inclusion of sufficient angular momentum can allow the profile 
to be arbitrarily shallow (depending on the initial profile). 
Their analytical model was restricted to
scale-free conditions and had an artificial prescription for adding
angular momentum. By construction their model led to power law final
profiles. In contrast the initial power spectrum we use is not scale-free; 
there is no imposed spherical symmetry; and the angular momentum is
acquired dynamically. Therefore we can only compare our profile to 
the predictions of infall models in a local, approximate sense. 

The comparison is nevertheless interesting because 
the slope of the NFW density profile at $r=a_n$
is $\alpha=-2$, precisely the value at which the secondary infall model
shows a transition from an 
infall profile dominated by radial motions, to one with significant
tangential motions. We have shown in figure~\ref{figevol} that $r=a_n$
does mark a transition in the relative importance of 
the ``centrifugal'' force due to 
the tangential velocity dispersion and the
radial component of the graviational force. The emergence of this 
transition coincides with that of an NFW-like profile with a slope
that is shallower than $-2$ in the inner regions, and both features
are stable over time once they appear at $z\sim 2-3$. 
Thus there are two qualitative changes in the density profile as 
haloes evolve from $z\sim 6$ to the present: the initial
steepening as mainly radial infall sets in, and the subsequent flattening of
the inner profile as tangential motions become significant. Both
features are in accord with infall models, and do not appear to be
disturbed by ongoing mergers. The same features are present 
in all the cluster haloes, and most significantly in the HDM
haloes which have much less merging but produce very similar
profiles. These qualitative features argue against the merger mechanism
proposed by Syer \& White (1996) to explain the formation of NFW-like
profiles. 

In the context of a violent relaxation scenario of halo formation,
the non-radial motions of halo particles would be induced by
mostly stochastic torques due to matter anisotropies. An 
alternative possibility is the radial orbit instability
(e.g. Carpintero \& Muzzio 1995) which exists even in the absence
of large scale fluctuations in the potential. It is important to
understand the origin of the torques that produce these tangential motions 
before they can be regarded as the dynamical agent
in forming the density profile. We plan to experiment with 
halo collapse simulations from varying 
initial conditions to establish if the mechanism
that produces an NFW-like profile is distinct from one in which 
mergers play the dominant role, though
clearly both processes operate in the evolution of dark matter haloes. 

\section{Conclusions}

We have investigated the formation and evolution of galaxy 
clusters in different
cosmogonies by means of numerical simulations. The investigated cosmogonies
include standard cold dark matter models ($\Omega=1$), open models
($\Omega=0.3$), models with a cosmological constant ($\Omega=0.3$,
$\Lambda=0.7$), 3 mixed dark matter models with differing $\Omega_\nu$ and
differing number of massive neutrino species as well as a pure hot dark matter
model ($\Omega=1$). 3 different realisations of each model have been
investigated.  The simulation have been performed using newly developed high
resolution N-body code which combines a PM code with a high resolution
tree code or a direct summation N-body code based on the special purpose
hardware GRAPE. High spatial and mass resolution is achieved by using an
enhanced multi-mass technique and a hierarchically nested particle
arrangement. Test simulations demonstrate that our chosen softening and time
stepping are consistent and that results have numerically converged.

In all considered cosmogonies, except the hot dark matter model, structure forms
due to hierarchical clustering.  For identical phases of the Gaussian random
field the evolution pattern of the halo formation is similar. Differences in the
cosmological parameters and the power spectra mainly manifest in a different
time evolution, but the merging pattern, like \eg number of mergers and mass
ratio of mergers is fairly similar. Consistent with the predictions of linear
theory, the evolution proceeds earlier in models with $\Omega < 1$, while in
mixed dark matter models the evolution is retarded to low redshifts.  As
expected the evolution pattern of the HDM clusters is different as it
is dominated by the
initially smooth distribution of particles.  The differences between
individual realisations of a given cosmological model are large and much
stronger than differences within differing cosmogonies for a given realisation.
Gross properties of clusters like size or angular momentum can be well described
by analytic models like,
\eg, the spherical top hat model.

At all considered redshifts, the density profile of the virialised
portion of a galaxy clusters can be well described by the 
NFW profile (Navarro et al. 1996a),
which has a slope of $\alpha=-1$ near the center and of $\alpha=-3$ at large
radii. The NFW profile provides a very good description for all considered
cosmogonies and the influence of the background cosmological model only
influences the central matter concentration and the characteristic radius. Due
to their earlier formation epoch, low $\Omega$ models are typically more
concentrated than high $\Omega$ CDM modes. Analogously, mixed dark matter models
are slightly less concentrated due to their retarded formation epoch. 
Our
results extend earlier findings to other cosmological models and supports the
interpretation that the NFW profile is generic for all hierarchically
clustering scenarios. It is
suggested that the characteristic radius $a_n$ and the shallow density profile
($\alpha > -2$) for $r<a_n$ are linked to the non-radial motions of
particles. 
The mechanism that produces the tangential velocity
dispersion of infalling particles and its possible dynamical role
in the evolution of the density profile merits further
investigation.

\section*{Acknowledgements}
 
We  wish to thank Dave Syer, Bepi Tormen and Simon White for many helpful 
discussions. This work was supported by the
Sonderforschungsbereich SFB 375-95 of the Deutsche
Forschungsgemeinschaft.

\end{document}